\documentclass{article}

\usepackage{amssymb,amsfonts,amsmath}
\usepackage{cite,enumerate,float,indentfirst}
\usepackage[dvips]{graphicx}
\usepackage{color}

\def\be{\begin{eqnarray}}
\def\ee{\end{eqnarray}}
\def\nn{\nonumber}

\def\Tr{{\rm Tr}\,}

\definecolor{red}{rgb}{1,0,0}
\definecolor{orange}{rgb}{1,0.5,0}
\definecolor{violet}{rgb}{0.7,0,1}

\hoffset= - 1.0in         

\begin{document}

\title{\vspace{.1cm}{\Large {\bf {Gaussian distribution of LMOV numbers}}\vspace{.2cm}}}
\author{
{\bf A.Mironov$^{a,b,c}$}\footnote{mironov@lpi.ru; mironov@itep.ru},
\ {\bf A.Morozov$^{b,c}$}\thanks{morozov@itep.ru},
\ {\bf An.Morozov$^{b,c,d}$}\footnote{andrey.morozov@itep.ru},
\ \ and
 \ {\bf A.Sleptsov$^{b,c,d}$}\thanks{sleptsov@itep.ru}}
\date{ }

\maketitle

\vspace{-5cm}

\begin{center}
\hfill FIAN/TD-11/17\\
\hfill IITP/TH-09/17\\
\hfill ITEP/TH-15/17
\end{center}

\vspace{2.3cm}

\begin{center}
$^a$ {\small {\it Lebedev Physics Institute, Moscow 119991, Russia}}\\
$^b$ {\small {\it ITEP, Moscow 117218, Russia}}\\
$^c$ {\small {\it Institute for Information Transmission Problems, Moscow 127994, Russia}}\\
$^d$ {\small {\it Laboratory of Quantum Topology, Chelyabinsk State University, Chelyabinsk 454001, Russia }}

\end{center}

\vspace{.5cm}

\begin{abstract}
Recent advances in knot polynomial calculus
allowed us to obtain a huge variety of LMOV integers
counting degeneracy of the BPS spectrum of topological theories
on the resolved conifold and
appearing in the genus expansion of the plethystic logarithm of the Ooguri-Vafa
partition functions.
Already the very first look at this data reveals
that the LMOV numbers are
randomly distributed in genus (!) and are very well
parameterized by just three parameters depending
on the representation, an integer and the knot.
We present an accurate formulation
and evidence in support of this new puzzling observation
about the old puzzling quantities.
It probably implies that the BPS states, counted by the
LMOV numbers can actually be composites made from some still
more elementary objects.
\end{abstract}

\bigskip

\section{Introduction}

The LMOV numbers $N^{\cal K}_{Q,g,n}$ \cite{OV,LMVm,LMV,LM} which are constructed from knot polynomials
are among the most mysterious objects in modern theoretical physics.
So far they were at the top of the hierarchy of integrality properties
in knot theory, though our new observations could imply that they are
still an intermediate stage.

Anyhow, the beginning of the story is integrality of
coefficients of the knot polynomials themselves, which
is quite mysterious from the point of view of the underlying Yang-Mills
(Chern-Simons) theory. Indeed, the knot polynomials are the Wilson loop averages in this theory \cite{Wit}, and, if expanded in powers of non-perturbative variables
$q$ and $A$, they turn out to possess integer coefficients.
Attempts to find a "physical reason" for this are still in progress
and they involve counting of states in some artificially constructed
higher-dimensional theories \cite{HD}.
Another artificial way to overcome the problem was found in using
alternative definitions of the knot polynomials: through
${\cal R}$-matrices, skein relations
(reflecting the Hecke algebra properties of the fundamental ${\cal R}$-matrices)
and building associated complexes, for which the knot polynomials are just
the Euler characteristics (and the Poincare polynomials provide even more
general integer-valued topological invariants, with no any Yang-Mills
interpretation yet).

Another integrality was observed in the Ooguri-Vafa partition
functions $Z_{OV}$, which include sums over HOMFLY polynomials \cite{knotpol,Con} over all
representations \cite{OV}.
The Ooguri-Vafa integers were then interpreted as those counting the
numbers of BPS states (or, more precisely, their charges) in some dual
theories (of topological strings on the resolved conifold).
However, it was soon realized that these numerous numbers are not
all algebraically independent, and this brought into the game
a further genus expansion $Z_{OV}$, which provided a decomposition
of the Ooguri-Vafa integers into sums of "more fundamental" LMOV numbers \cite{LMVm,LMV,LM}.
No algebraic relations were so far found among the LMOV numbers,
and they are often considered as counting the number of "independent
BPS states".

The LMOV numbers characterize the genus expansion of the Hurwitz transform of
the plethystic logarithm of the Ooguri-Vafa
partition function for the knot ${\cal K}$ (not to be confused with the genus expansion of the Ooguri-Vafa partition function itself \cite{MMS}),
i.e. are defined by a somewhat complicated chain of transformations
from the colored HOMFLY-PT polynomials:
\be
\sum_R H_R^{\cal K}(A,q)\chi_{_R}\{\bar p\}
= \exp\left\{\sum_{d=1}^\infty \frac{1}{d}\,\widehat{\rm Ad}_d
\underbrace{\left(
\sum_{Q,g,n } N^{\cal K}_{Q,g,n} A^n (q-q^{-1})^{2g-2}\chi_{_Q}\{\bar p^{\vee}\}
\right)
}_{\sum_{Q,g,n } \tilde N^{\cal K}_{Q,k,n} A^n q^k \chi_{_Q}\{\bar p\}}
\right\}
\label{OVfrE}
\ee
Here the (unreduced) HOMFLY-PT polynomials
$H_R^{\cal K}(A,q) =
\left< \Tr_R P\exp \oint_{\cal K}{\cal A}\right>_{CS}^{SL(N)}$
at the l.h.s are
Laurent polynomials in $A$ and $q$ with integer coefficients,
"colored" by the Young diagrams $R$,
the average is taken in the Chern-Simons theory with the gauge $SU(N)$
and the coupling constant is $\kappa$, then, $q=\exp\frac{2\pi i}{\kappa+N}$ and $A=q^N$.
$\chi_{_Q}\{\bar p\}$ are the characters of the linear group (Schur polynomials
of the "time-variables" $\bar p_m$), where $Q$ is the Young diagram
$Q=\{Q_1\geq Q_2\geq\ldots\geq Q_{l_{Q}}>0\}$, and the shifted time variables are
\be\label{psh}
\bar p_k^\vee=(q^k-q^{-k})\, \bar p_k
\label{newbarp}
\ee
The generalized Adams transform $\widehat{\rm Ad}_d$
raises each $A$ and $q$ to the power $d$
and scales all time variable modes: $\bar p_m \longrightarrow \bar p_{md}$.
The pole in $z\equiv q-q^{-1}=0$ at the r.h.s. of (\ref{OVfrE}) is actually
of the first order because the empty Young diagram $Q=\emptyset$ does not contribute and any other character contains at least one factor
$(q-q^{-1})$ coming from shifted time variables.
This expansion has certain parallels with the standard genus expansions
(though not very precise yet), thus we denote the corresponding summation
variable $g$ and call it "genus".\footnote{
For a much more accurate notion of genus expansion for the particular HOMFLY-PT
polynomials and its interesting properties see \cite{MMS},
but its relation to the $g$-expansion of the Ooguri-Vafa plethystic free energy
(\ref{OVfrE}) is somewhat obscure.}

The LMOV formula in the main line of (\ref{OVfrE})
differs from the earlier Ooguri-Vafa (OV) expansion \cite{OV}, written beneath it,
by rescaling (\ref{newbarp}) of auxiliary time-variables $\bar p_k$.
Because of the general transposition properties of characters
$\chi_{_{R^{\rm tr}}}\{\bar p_k\}=\chi_{_R}\left\{(-)^{k+1}\bar p_k\right\}$
and the HOMFLY polynomials
$H_{R^{\rm tr}}(q)=H_{R}(-1/q)$, this rescaling
makes the remaining coefficients invariant under the change $q\longrightarrow -q$
and they can re-expressed through the Toda-like
(logarithmic) variable $z=q-q^{-1}$.
In result, the algebraically-dependent OV numbers $\hat N$ are decomposed into independent LMOV integers $N$.
Moreover, in variance with the chaotically-looking set of $\hat N$, that of $N$
acquires a pronounced structure, which will be in the focus of the present Letter.

As already mentioned, $N^{\cal K}_{Q,g,n}$  are all {\it integers}
and there are finitely many of them for each given $Q$ and $n$,
which is an implication of the hypothetical Gopakumar-Vafa duality
and its further refinements \cite{GV}.
Testing this integrality property is a biggest challenge for any
efficient approach to knot calculations, and only few results were
available on this.
Recently, as an application of a new spectacular breakthrough \cite{arbor,evo3,MMMRS,MMf,Rama2} with
arborescent knot calculus \cite{Con,Cau},
we performed a new attack on the problem \cite{Rama3}, which provided quite
large lists of the LMOV numbers so that now their properties can be
analyzed in reasonable detail (see also some other attempts in \cite{Zhu1}).

First of all, their integrality and finiteness were confirmed once again, now with a really big set of data.

Second, new features showed up, one of which we describe in the present letter:
{\bf with a very high accuracy the LMOV numbers appeared to be distributed
by a Gaussian law} as functions of $g$.
In this paper we do not go into discussion of the {\it reasons} for this phenomenon,
our goal at this stage is to present {\it an evidence} in support of
this claim and {\it to measure} parameters of the distributions as
functions of representations and knots.

\begin{figure}[t]
\includegraphics[bb = 0 0 14cm 13cm,width=150pt,height=150pt]{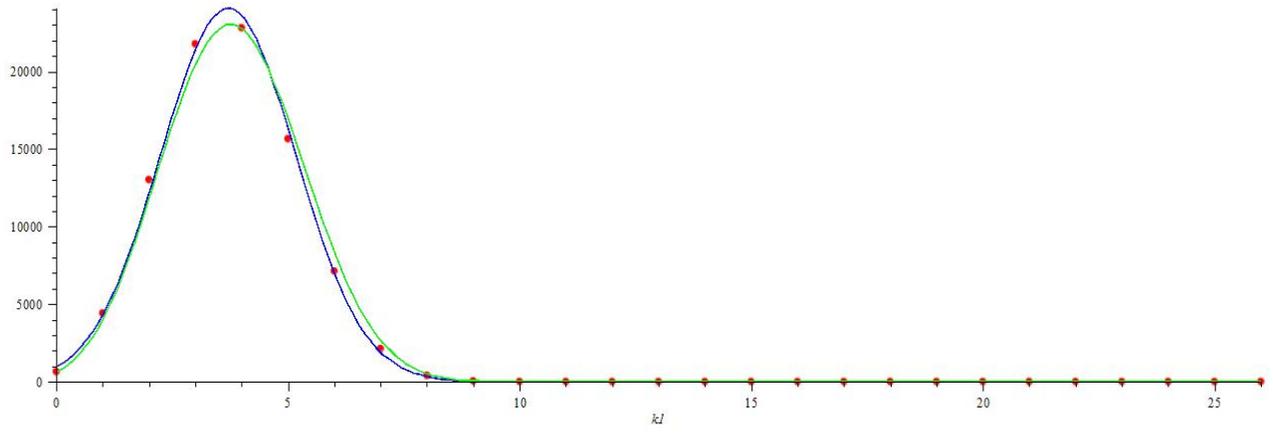}
\caption{\footnotesize
{\bf Red points:} The LMOV numbers $N_{[4],g,0}^{8_{20}}$ plotted as a function of $g$,
the number of points is $27$ (power in $z^2=(q-q^{-1})^2$ is $26$).
{\bf Blue line:} The Gaussian curve (\ref{Gaussfit}) with parameters
$\mu=3.71$, $\sigma=1.46$, $I=88053$.
{\bf Green line:} The binomial curve $ I\cdot  C^N_g\cdot p^g(1-p)^{N-g}$
with $I=88053$, $N=20$ and $p=0.53$.
}
\label{mainexa}
\end{figure}

Thus, it turns out that, with a very good accuracy for any given knot ${\cal K}$,
the representation $Q$ and the parameter $n$, the LMOV numbers as a function of
remaining variable $g$ are described by the formula (its accuracy being illustrated in Fig.\ref{mainexa})

\be
\boxed{
N^{\cal K}_{Q,g,n} \approx
G^{\cal K}_{Q,n}(g) \equiv
(\pm 1)\cdot\frac{I_{Q,n}^{\cal K} }{\sqrt{2\pi}\,\sigma_{Q,n}^{\cal K} }
\exp\left( -\frac{(g-\mu_{Q,n}^{\cal K} )^2}{2\sigma_{Q,n}^2}\right)
}
\label{Gaussfit}
\ee
with only three parameters $\mu$, $\sigma$ and $I$
which can depend on ${\cal K}$, $Q$ and $n$, but not on the genus $g$. Note that in most cases the signs of the LMOV numbers are the same in the whole column (only at the very ends they sometimes happen to be different, where they are very small as compared with typical numbers in the column), which one notable exception that we discuss in s.\ref{3.3}. Thus, for the sake of uniformity, in all plots in the paper we choose the sign in such a way that the maximal LMOV numbers in the column are positive (see the issue of signs in \cite{Naw}).

Figure 1 is a visualization of one particular column from one particular table
in \cite{Rama3}, $\mathbf{N}_{[ 4]}:$

\bigskip

\hspace{-2cm}{\tiny
\begin{tabular}{|c|ccccccccccccc|}
\hline
&&&&&&&&&&&&&\\
$ g \backslash n=$ & -20 & -18 & -16 & -14 & -12 & -10 & -8 & -6 & -4 &
-2 & 0 & 2 & 4 \\
&&&&&&&&&&&&&\\
\hline
&&&&&&&&&&&&&\\
0 & 11440 & -87173 & 293893 & -576270 & 726572 & -614639 & 352840 &
-135087 & 31946 & -3116 & -645 & 269 & -30 \\
&&&&&&&&&&&&&\\
1 & 228250 & -1635276 & 5137191 & -9286702 & 10657519 & -8081601 &
4086664 & -1355188 & 275713 & -23835 & -4413 & 1934 & -256 \\
&&&&&&&&&&&&&\\
2 & 2386083 & -16136564 & 47369508 & -79062965 & 82549940 & -55906878 &
24662653 & -6936301 & 1167700 & -85260 & -13024 & 6054 & -946 \\
&&&&&&&&&&&&&\\
3 & 16661172 & -107057436 & 295234625 & -456862001 & 435207944 &
-263432562 & 101033685 & -23791571 & 3205258 & -186183 & -21779 & 10693
& -1845 \\
&&&&&&&&&&&&&\\
4 & 84507887 & -519907050 & 1355251916 & -1954018736 & 1704701161 &
-924387333 & 308130762 & -60328166 & 6336339 & -273569 & -22825 & 11672
& -2058 \\
&&&&&&&&&&&&&\\
5 & 324218115 & -1925277128 & 4773608162 & -6443894762 & 5167017105 &
-2515811031 & 729145727 & -118202574 & 9488612 & -283394 & -15660 & 8205
& -1377 \\
&&&&&&&&&&&&&\\
6 & 964060168 & -5570786966 & 13216800689 & -16777389869 & 12401705555 &
-5430400831 & 1368154785 & -182973872 & 11046274 & -212013 & -7130 &
3771 & -561 \\
&&&&&&&&&&&&&\\
7 & 2260162822 & -12811019508 & 29245594272 & -35043455568 & 23932680078
& -9429839738 & 2062360089 & -226492258 & 10126690 & -115736 & -2131 &
1124 & -136 \\
&&&&&&&&&&&&&\\
8 & 4233247530 & -23722411643 & 52372794915 & -59428427145 & 37551473309
& -13307457043 & 2519220675 & -225742273 & 7347950 & -46065 & -401 & 209
& -18 \\
&&&&&&&&&&&&&\\
9 & 6401577363 & -35742184785 & 76665320902 & -82599999575 & 48322505041
& -15378284326 & 2508707131 & -181847621 & 4219095 & -13203 & -43 & 22 &
-1 \\
&&&&&&&&&&&&&\\
10 & 7882325987 & -44189952927 & 92475676765 & -94805633387 &
51345839966 & -14636437277 & 2044843043 & -118567832 & 1908313 & -2650 &
-2 & 1 & 0 \\
&&&&&&&&&&&&&\\
11 & 7955332198 & -45135236322 & 92502848394 & -90392118364 &
45284233622 & -11520499158 & 1367294612 & -62528465 & 673836 & -353 & 0
& 0 & 0 \\
&&&&&&&&&&&&&\\
12 & 6613694775 & -38282967105 & 77104986495 & -71912996672 &
33272091742 & -7518694886 & 750290984 & -26588381 & 183076 & -28 & 0 & 0
& 0 \\
&&&&&&&&&&&&&\\
13 & 4544009056 & -27063745212 & 53736998578 & -47883973931 &
20411212868 & -4072813151 & 337335831 & -9061459 & 37421 & -1 & 0 & 0 & 0 \\
&&&&&&&&&&&&&\\
14 & 2584289462 & -15981116400 & 31372940707 & -26729420170 &
10461927572 & -1829952056 & 123776261 & -2450930 & 5554 & 0 & 0 & 0 & 0 \\
&&&&&&&&&&&&&\\
15 & 1216390788 & -7887415145 & 15349485952 & -12510461549 & 4475947275
& -680239806 & 36810141 & -518220 & 564 & 0 & 0 & 0 & 0 \\
&&&&&&&&&&&&&\\
16 & 472819775 & -3250240468 & 6285479701 & -4902402599 & 1593840148 &
-208191240 & 8778345 & -83697 & 35 & 0 & 0 & 0 & 0 \\
&&&&&&&&&&&&&\\
17 & 151097941 & -1114995248 & 2147461941 & -1603116813 & 469971253 &
-52061194 & 1652078 & -9959 & 1 & 0 & 0 & 0 & 0 \\
&&&&&&&&&&&&&\\
18 & 39405913 & -316778506 & 608869085 & -435066095 & 113847320 &
-10516465 & 239570 & -822 & 0 & 0 & 0 & 0 & 0 \\
&&&&&&&&&&&&&\\
19 & 8294017 & -73942242 & 142098694 & -97183015 & 22394545 & -1687755 &
25798 & -42 & 0 & 0 & 0 & 0 & 0 \\
&&&&&&&&&&&&&\\
20 & 1385933 & -14015257 & 26975820 & -17655709 & 3517260 & -209987 &
1941 & -1 & 0 & 0 & 0 & 0 & 0 \\
&&&&&&&&&&&&&\\
21 & 179446 & -2120970 & 4095152 & -2564506 & 430297 & -19510 & 91 & 0 &
0 & 0 & 0 & 0 & 0 \\
&&&&&&&&&&&&&\\
22 & 17344 & -249997 & 484911 & -290476 & 39489 & -1273 & 2 & 0 & 0 & 0
& 0 & 0 & 0 \\
&&&&&&&&&&&&&\\
23 & 1177 & -22102 & 43125 & -24704 & 2556 & -52 & 0 & 0 & 0 & 0 & 0 & 0
& 0 \\
&&&&&&&&&&&&&\\
24 & 50 & -1378 & 2708 & -1483 & 104 & -1 & 0 & 0 & 0 & 0 & 0 & 0 & 0 \\
&&&&&&&&&&&&&\\
25 & 1 & -54 & 107 & -56 & 2 & 0 & 0 & 0 & 0 & 0 & 0 & 0 & 0 \\
&&&&&&&&&&&&&\\
26 & 0 & -1 & 2 & -1 & 0 & 0 & 0 & 0 & 0 & 0 & 0 & 0 & 0 \\
&&&&&&&&&&&&&\\
\hline
\end{tabular}
}

\bigskip

\bigskip

\begin{figure}[b]
\bigskip
\includegraphics[bb = 0 0 10cm 13cm,width=150pt,height=150pt]{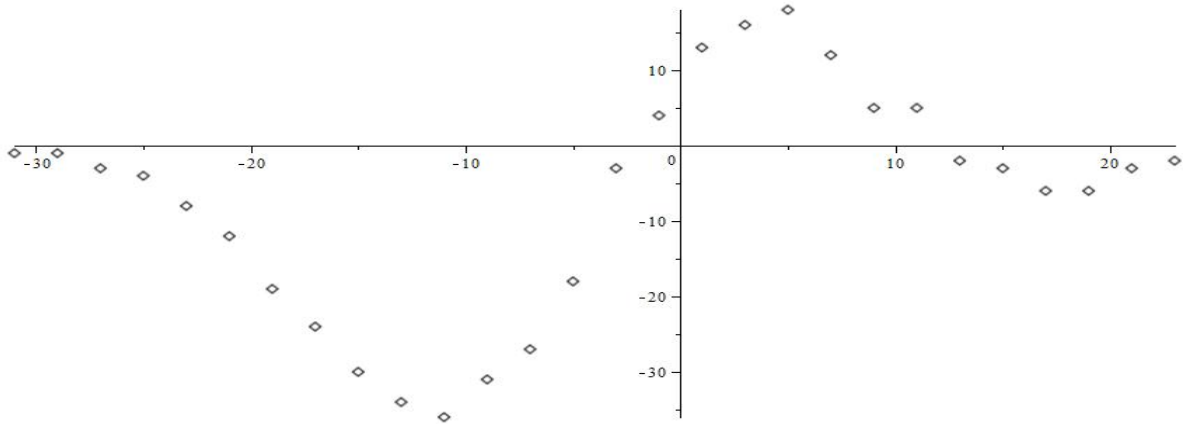}
\caption{\footnotesize
Original Ooguri-numbers $\tilde N^{8_{20}}_{[4],k,0}$ plotted as a function of $k$.
Note that these are coefficients in front of $q^k$, not $z^{2g}=(q-q^{-1})^{2g}$
as in Fig.\ref{mainexa}. Also the OV polynomial for representation $\Delta=[4]$
is a sum of LMOV polynomials over five representations $[4],[31],[22],[211],[1111]$ --
thus there is no way to directly compare this picture with Fig.\ref{mainexa}.
It only illustrates the fact that {\it not any} relevant polynomial
has Gaussian distribution of coefficients.
}
\label{OVexa}
\end{figure}

\noindent
For a plot of the first column from this table see \cite[Fig.1]{Rama3},
for more plots, s.\ref{results} in the present paper.
For a comparison, we show in Fig.\ref{OVexa}
a similar plot for the Ooguri-Vafa numbers $\tilde N_{\Delta,k,j}$.

If true, this observation
could make a hint that the LMOV numbers are not {\it fundamental},
and their variety is nearly $3_R$-dimensional for every $Q$ and $n$:
\be
\{N_{Q,g,n}\} \ \longrightarrow \ \{\mu_{Q,n}, I_{Q,n}, \sigma_{Q,n}\}
\ee
In fact, the main natural source of Gaussian functions
is the binomial distribution $C^{2N}_{N+m} p^m(1-p)^{N-m}$ at large $N$,
i.e. the emergence of the Gaussian distribution could be a clear sign
that the LMOV numbers count not the "primary" BPS states,
but "composites" of some many more elementary ones, this is what
makes them randomly distributed.
An identification and, to begin with, {\it counting} of these elementary
excitations is the next immediate task of analysis of our data.

Of course, this type of compositeness does not make the LMOV numbers
{\it algebraically} dependent, but definitely makes them
unrelated to the "fundamental degrees of freedom".

The fact that some non-vanishing genera are distinguished
(that distributions are located around non-vanishing $\mu$)
can signal that there can be soliton- or instanton-like configurations,
perhaps, winding non-trivially around the knot.
Then, for entropy reasons, non-vanishing windings can dominate
in the quasiclassical approximation, thus, making the genus expansion
peaked around non-vanishing genera.

Whatever the reason, this peaking
looks like a new and very interesting phenomenon in the emerging
theory of the genus expansions and Hurwitz tau-functions.
Among immediate questions is what are associated peculiarities
of the spectral curves and AMM/EO topological recursion.

To complete these introductory remarks, let us make a few comments. First of all, the Gaussian curve is a limiting case of the binomial distribution at large enough $N$. In fact, $N$ should not be too large, and one can see in Fig.\ref{mainexa} that one can fit the distribution of the LMOV numbers by the binomial distribution as well, with $N$ for the distribution being equal to 20. Hence, one may equally well think of the LMOV number distributions in binomial terms.

Second, with increase of the representation, the number of LMOV numbers also increases. One can easily see that the approximation to the Gaussian curve also becomes better with this increase. Hence, one could expect that the deviations from the Gaussian curve can be due to too little number of points and too small their numerical values, both increasing with the representation. Hence, the conjecture ultimately sounds:

\paragraph{Conjecture.}
{\it The distribution of the LMOV numbers at different genera is described by the Gaussian curve (\ref{Gaussfit}) asymptotically at large representations with all three parameters $\mu$, $\sigma$ and $I$ increasing with the representation.}

\bigskip

Though, as we shall see in examples below this distribution is typically very close to the Gaussian curve at very first representations.

\bigskip

In this letter, we present some typical pieces of the data,
in the form of plots and
also tables, which describe the dependence of parameters $\mu$, $\sigma$
and $A$ on their arguments.
Among interesting general properties is the increase of dispersion
$\sigma$ with the increase of the minimal number of strands needed
to describe the knot.

We also provide some initial results about the distribution of LMOV numbers
in $n$,
which may look more like a superposition of Gaussian distributions which are better separated
in well-structured knots and overlap to turn into a single Gaussian
(i.e. the numbers get fully uncorrelated) for the knots without a visible structure.
This demonstrates the ability to use the LMOV number distributions for
quantitative description of the "true complexity" of knots,
the option long looked for and not yet provided by any other type of
knot invariants.

\section{Theoretical background}

\subsection{Counting states}

The partition function usually counts the number of exited states ${\cal N}_E$
as a function of the excitation level $E$:
\be
{\cal Z}(q) = \sum_E {\cal N}_E\cdot q^E
\ee
(one can of course add other characteristics of the microcanonical ensembles
and get a function of several variables, not just the temperature $q=e^{-1/T}$).
In quantum field theory, these states are elements of
the Fock space, which is huge even for a {\it single} particle.
This happens for two reasons: a particle can be treated as
a collection of oscillators, and there are many oscillators
(as many as momenta) and each oscillator has many states.
The first kind of multiplicity is put under control when the space
is compactified: then, the momenta become discrete and enumerable,
and in the state counting we actually treat the fields with different
discrete momenta as different particles (in particle
phenomenology, this is because different momenta in compact dimensions
provide different particles in $4d$,
in pure string theory considerations, {\it all} dimensions
are compactified to make {\it all} momenta discrete).
The second multiplicity is eliminated if we consider the number
of oscillators instead of the number of states, i.e.
count the states in the Hilbert space underlying the Fock space.
Technically, this corresponds to replacing the partition function with
its plethystic logarithm (see s.\ref{plexp} below)
\be
{\cal Z}(q) = {\rm Plexp}\Big({\cal F}(q)\Big)
\ee
where
\be
{\cal F}(q) = \sum_\omega {\cal M}_\omega q^\omega
\ee
and ${\cal M}_\omega$ is a number of {\it oscillators} with frequency $\omega$.

In string theory, each string field contains a well-ordered tower of oscillators,
thus, the further appropriate modification of the free energy is
\be
 {\cal F}(q) = \sum_\omega M_\omega \frac{q^\omega}{1-q^{2\omega}}
 = -\sum_\omega \frac{M_\omega}{q^\omega-q^{-\omega}}
\label{strefcount}
\ee
where $M_\omega$ is now the number of string fields, i.e. of oscillator {\it towers}.
For branes, the towers are bigger, then higher is also
the pole in the denominator.
For example, for the MacMahon modules (and/or with representations of DIM algebras, see \cite{DIM}),
we would get
\be
 {\cal F}_{McM}(q) =\sum_\omega \frac{m_\omega}{(q^{\omega}-q^{-\omega})^2}
\ee
where $m_\omega$ is the number of MacMahon modules.

LMOV numbers parameterize the set $\{M_\omega\}$, i.e. the spectrum of $\omega$
for some particular string models, associated with knots.
Since all stringy oscillator towers are already included into the
{\it definition} of $M_\omega$, these numbers can be considered as
counting the states in a {\it topological} string theory,
i.e. the elements of the underlying chiral ring.
Actually counted are bundles over $D2$-branes of genus $g$, thus the "primary" numbers
carry this index.
There are also $l$ holes, what gives rise to Chan-Paton factors and additional
resummation over representations $Q$ of symmetric group $S_l$.
In result  $M_\omega$ in (\ref{strefcount}) in the case Chern-Simons theory
are decomposed into $N_{Q,g,n}$, and the rescaling of times
in (\ref{psh}) are taking into account the knot-independent enumeration
of bundles and boundary conditions, see \cite{LMVm} for details.

Before going back to interpretation of our results from the point of view
of the state counting, we remind two pieces from textbooks:
the basics of plethystic calculus and binomial origins of the Gaussian distributions.

\subsection{Plethystic exponentials
\label{plexp}}

For a function $f(A,q\,|\,p_n,\bar p_n)$ of $A$, $q$ and time variables $\{p_n\}$, $\{\bar p_n\}$,
plethystic exponential is defined as
\be
{\rm Plexp}(f) =
\exp\left\{\sum_{d=1}^\infty \frac{1}{d}\,\widehat{{\rm Ad}}_d\big(f\big)\right\}
\equiv
\exp\left(\sum_{d=1}^\infty \frac{f(A^d,q^d\,|\,p_{nd},\bar p_{nd})}{d} \right)
\ee
Thus, it incorporates the conventional
Adams operation $\widehat{{\rm Ad}}_d (p_n) = p_{nd}$
and preserves the topological locus \cite{DMMSS,MMM}
$p^*_n = \frac{A^n-A^{-n}}{q^n-q^{-n}}$
by changing all the ingredients of this formula in the same way.
It also inherits the main property of the ordinary exponential:
\be
{\rm Plexp}(f_1+f_2) = {\rm Plexp}(f_1)\cdot {\rm Plexp}(f_2)
\ee
The two most important examples of plethystic exponentials are
\be
\frac{1}{1-q} = {\rm Plexp}(q)
\ee
what provides the simple spectrum counting rules
\be
\prod_{i}^\infty \frac{1}{1-q^{k_i}} = {\rm Plexp}\left(\sum_i q^{k_i}\right)
\ee
like
\be
\prod_{k=1}^\infty \frac{1}{1-q^{k}} = {\rm Plexp}\left(\sum_k q^k\right)
={\rm Plexp}\left(\frac{q}{1-q}\right)
\ee
or
\be
\prod_{k=1}^\infty \frac{1}{(1-q^{k})^k} = {\rm Plexp}\left(\sum_k kq^k\right)
={\rm Plexp}\left(\frac{q}{(1-q)^2}\right)
\ee
and
\be
\sum_{R} \chi_{_R}\{p\}\chi_{_R}\{\bar p\} = \exp\left(\sum_k \frac{p_k\bar p_k}{k}\right)
= {\rm Plexp}\Big(p_1\bar p_1\Big)
\ee
As a consequence, the Ooguri-Vafa function for the unknot is
\be
Z_{\rm OV}^{\rm unknot}\{\bar p\} = \sum_R D_R(A,q)\chi_{_R}\{\bar p\} =
{\rm Plexp}\Big( p_1^*\bar p_1\Big) =
{\rm Plexp}\left(\frac{A-A^{-1}}{q-q^{-1}}\cdot\bar p_1\right)
\ee

From the special polynomial \cite{DMMSS,IMMMfe,Zhu}, which is the specialization of the unreduced HOMFLY-PT polynomial at $q=1$:
$H_R^{\cal K}(A,q=1) =D_R\cdot \Big(\sigma^{\cal K}(A)\Big)^{|R|}$,
one can construct a quantity
\be
{\rm Plexp}\Big(\frac{A-A^{-1}}{q-q^{-1}} \cdot\bar p_1 \sigma^{\cal K}(A)\Big) =
\exp\left(\sum_{k=1}^{\infty} \frac{p_k^*\bar p_k}{k}\cdot\sigma^{\cal K}(A^k)\right)
\ee
which is different from the leading asymptotic of the Ooguri-Vafa function at $q=1$,
\be
{\rm log}\Big(Z_{\rm OV}^{\cal K}(A,q)\Big) \sim
{\rm log}\left(\sum_R  D_R(A|q) \cdot\Big(\sigma^{\cal K}(A)\Big)^{|R|}
\cdot\chi_{_R}\{\bar p\}\right)
= \sum_{k=1} \frac{p_k^*\bar p_k}{k}\cdot \Big(\sigma^{\cal K}(A)\Big)^k
\ee
because $\Big(\sigma^{\cal K}(A)\Big)^k \neq \sigma^{\cal K}(A^k)$. This explains why the genus expansion of the plethystic logarithm of the Ooguri-Vafa partition function differs from the genus expansion of the Ooguri-Vafa partition function itself \cite{MMS}.

\subsection{Symmetric group calculus}

To make our presentation technically complete we briefly repeat now some simple
formulas from \cite{LMVm}.
The OV and LMOV numbers are counting the maps of Riemann surfaces with boundaries
(interpreted as BPS $D2$-branes ending on $M$ $D4$-branes wrapping a Lagrangian submanifold ${\cal L}_K$ in $T^*S^3$ which passes through the given knot or link ${\cal K}\subset S^3$ so that ${\cal L}_K$ is the conormal bundle of ${\cal K}$).
According to \cite{LMVm}, actually counted are elements of
\be
{\rm Sym}\left\{F^n\otimes H^*\Big((S^1)^{2g+n-1}\Big)\otimes H^*({\cal M}_{g,n})\right\}
\ee
where $g$ and $n$ are the genus and number of holes, ${\cal M}_{g,n}$ is the
moduli space of maps, while additional two factors are cohomologies of the Jacobian
and the Chan-Paton factors at the boundary, $F$ stands for the fundamental
representation of auxiliary $SU(M)$ with monodromies parameterized by the
auxiliary times $\bar p_k$.
The symmetrization is over the $n$ boundary components and it is handled with the
help of the general rule:
\be\label{dec}
{\rm Sym}\Big\{A_{i_1,\ldots,i_n}\otimes B_{i_1,\ldots,i_n}\otimes
C_{i_1,\ldots, i_n}\Big\} \equiv
\sum_{P\in S_n} A_{P(i_1,\ldots,i_n)}\otimes B_{P(i_1,\ldots,i_n)}\otimes
C_{P(i_1,\ldots, i_n)} =
\sum_{R',R''} A_{R'\circ R'' }\otimes
B_{R' }\otimes C_{R'' }
\ee
where the sum at the l.h.s. goes over all the permutations from $S_n$,
while that at the r.h.s., over all Young diagrams $R'$ and $R''$ of the size $n$ labelling the irreps of $S_n$,
their product being denoted by $\circ$.
The meaning of this simple combinatorial formula
should be clear from a couple of examples at $n=2$:
\be\label{S1}
\frac{1}{2}\left(A_{1,2}\otimes B_{1,2} + A_{2,1}\otimes B_{2,1}\right) =
\frac{1}{2}\left(A_{1,2}+A_{2,1}\right)\otimes \frac{1}{2}\left(B_{1,2} + B_{2,1}\right)
+ \frac{1}{2}\left(A_{1,2}-A_{2,1}\right)\otimes \frac{1}{2}\left(B_{1,2} - B_{2,1}\right)
= \nn \\
= A_{[2]}\otimes B_{[2]} + A_{[1,1]}\otimes B_{[1,1]} \ \ \ \ \ \ \
\ee
and
\be
\frac{1}{2}\left(A_{1,2}\otimes B_{1,2}\otimes C_{1,2} + A_{2,1}\otimes B_{2,1}\otimes C_{2,1}\right) = \nn \\
= \frac{1}{2}\left(A_{1,2}+A_{2,1}\right)\otimes \frac{1}{2}\left(B_{1,2} + B_{2,1}\right)
\otimes \frac{1}{2}\left(C_{1,2} + C_{2,1}\right) +
\frac{1}{2}\left(A_{1,2}+A_{2,1}\right)\otimes \frac{1}{2}\left(B_{1,2} - B_{2,1}\right)
\otimes \frac{1}{2}\left(C_{1,2} - C_{2,1}\right) +  \nn \\
+\frac{1}{2}\left(A_{1,2}-A_{2,1}\right)\otimes \frac{1}{2}\left(B_{1,2} + B_{2,1}\right)
\otimes \frac{1}{2}\left(C_{1,2} - C_{2,1}\right)
+ \frac{1}{2}\left(A_{1,2}-A_{2,1}\right)\otimes \frac{1}{2}\left(B_{1,2} - B_{2,1}\right)
\otimes \frac{1}{2}\left(C_{1,2} + C_{2,1}\right)  =
\nn
\ee
\be
= A_{[2]}\otimes B_{[2]}\otimes C_{[2]} + A_{[2]}\otimes B_{[1,1]}\otimes C_{[1,1]} +
A_{[1,1]}\otimes B_{[2]}\otimes C_{[1,1]} + A_{[1,1]}\otimes B_{[1,1]}\otimes C_{[2]}
\label{ABCexpan2}
\ee
associated with the composition rules (Clebsh-Gordon coefficients) $[2]\circ[2]=[1,1]\circ[1,1] = [2]$,
$[2]\circ [1,1]=[1,1]\circ [2]=[1,1]$.

Generally, one can rewrite (\ref{dec}) in the form
\be\label{dec1}
{\rm Sym}\Big\{A_{i_1,\ldots,i_n}\otimes B_{i_1,\ldots,i_n}\otimes
C_{i_1,\ldots, i_n}\Big\}=\sum_{R_1,R_2,R_3}C_{_{R_1R_2R_3}}\ \mathbb{S}_{_{R_1}}A\otimes\mathbb{S}_{_{R_2}}B\otimes\mathbb{S}_{_{R_3}}C
\ee
where $\mathbb{S}_{_R}$ denotes the Schur functor \cite{Fulton2,Fulton,Martin} and the Clebsh-Gordon coefficients $C_{_{R_1R_2R_3}}$ are manifestly given by the formula
\be
C_{_{R_1R_2R_3}}=\sum_\Delta {\psi_{_{R_1}}(\Delta)\psi_{_{R_2}}(\Delta)\psi_{_{R_3}}(\Delta)\over z_{_\Delta}}
\ee
Here $\psi_Q(\Delta)$ are the symmetric group characters, $\Delta=\{\delta_1\geq\delta_2\geq\ldots\geq \delta_{l_{\Delta}}>0\}$ is the Young diagram and $z_\Delta$ is the standard symmetric factor of the Young diagram (order of the automorphism) \cite{Fulton}.

Formula (\ref{dec}) is an immediate corollary of the formula (which particular example is given in (\ref{S1}))
\be
{\rm Sym}\Big(V_1\otimes V_2\Big)=\sum_R \mathbb{S}_{_R}V_1\otimes\mathbb{S}_{_R}V_2
\ee
and of the action of the Schur functor on the tensor product
\be\label{S2}
\mathbb{S}_{_R}\Big(V_1\otimes V_2\Big)=\sum_{R_1,R_2}C_{_{R_1R_2R_3}}\ \mathbb{S}_{_{R_1}}V_1\otimes\mathbb{S}_{_{R_2}}V_2
\ee
Using (\ref{S2}), one can further increase the number of factors in the tensor product (\ref{dec1}).

The next step is to convert the spaces into generating functions.
This includes converting the number $n$ with degrees of a parameter $A$, spins from $H^*(S^1)$,
with powers of $\pm q^{\rm \, spin}$
and representations from $F^{\otimes n}$, with $\bar p$-dependent characters.
Moreover, the contribution of circle cohomologies (from the $B$-factors)
can be absorbed into a redefinition of the characters:
for example, the last formula in (\ref{ABCexpan2}) can be rewritten as
\be
 \Big(B_{[2]}C_{[2]} + B_{[1,1]} C_{[1,1]} \Big)\,\chi_{[2]}\{\bar p\}
+ \Big(B_{[2]}C_{[1,1]} + B_{[1,1]}C_{[2]}\Big) \chi_{[1,1]}\{\bar p\}
=  C_{[2]}\,\chi_{[2]}^\vee\{\bar p\} +  C_{[1,1]}\,\chi_{[1,1]}^\vee\{\bar p\}
\ee
where
\be
\chi_{[2]}^\vee = B_{[2]}\,\chi_{[2]} + B_{[1,1]}\, \chi_{[1,1]}  \nn \\
\chi_{[1,1]}^\vee = B_{[1,1]}\, \chi_{[2]}+ B_{[2]}\,\chi_{[1,1]}
\label{cheveetrans}
\ee
In general, the OV partition function
is expected on the base of the open-closed-string
duality arguments  of \cite{GV,OV,LMVm,LMV,LM} to be:
\be
\boxed{
Z_{OV}^{\cal K} = \sum_R  H_R^{\cal K}(A,q)\,\chi_R\{\bar p\}
= {\rm Plexp}\left(
\frac{1}{q-q^{-1}}
\sum_{Q,g,n}  C_{_Q}^{\cal K}(g,n)\cdot (q-q^{-1})^{2g}\, A^n\, \chi_{_Q}^\vee\{\bar p\}
\right)
}
\ee
where the quantities
\be
C_{_Q}^{\cal K}(g,n)= {\rm Euler\ char}\left\{\mathbb{S}_{_R}
\Big(H^*({\cal M}_{g,n}^{\cal K})\Big)\right\} =
N_{Q,g,n}^{\cal K}
\ee
are exactly the LMOV numbers,
$(q-q^{-1})^{2g}$ is the power of $n$-independent contribution $q-q^{-1}$ from
$\Big(H^*(S^1)\Big)^{2g}$,
and the above example (\ref{cheveetrans}) shows that the only thing to calculate  are the
quantities $B_{R}$, which define the transformation from $\chi\{\bar p\}$
to $\chi^\vee\{\bar p\}$.
These $B$'s are cohomologies of the remaining $(S^1)^{n-1}$
weighted with the factors $\pm q^{\rm \,spin}$,
and calculations are very simple.
In the notation of \cite{LMVm}, if the non-trivial 1-form on $S^1$ is denoted by $\psi$,
then the basis in $H^*\Big((S^1)^{n-1}\Big)$ is made from the external products
$\psi_{i_1}\wedge \ldots \wedge \psi_{i_k}$ subject to the additional constraint
$\sum_{i=1}^n \psi_i = 0$.
Thus, the only non-vanishing are $B_R$ with single-hook Young diagrams $R$, and
\be
B_{[r,1^{n-r}]} = (-)^r q^{2r-n-1}
\ee
In particular, $B_{[2]}=q$, $B_{[1,1]}=-q^{-1}$
and
\be
\chi^\vee_{[2]}\{\bar p\} = q\chi_2\{\bar p\} -q^{-1}\chi_{[1,1]}\{\bar p\} =
\frac{1}{q-q^{-1}} \chi_{[2]}\{\bar p^\vee\}, \nn \\
\chi^\vee_{[1,1]}\{\bar p\} = -q^{-1}\chi_2\{\bar p\} +q\chi_{[1,1]}\{\bar p\} =
\frac{1}{q-q^{-1}} \chi_{[1,1]}\{\bar p^\vee\}
\ee
with
\be
\bar p_k^\vee = (q^k-q^{-k})\bar p_k,
\ee
Eq.(\ref{newbarp}) is a generalization of this simple example.

Technically, conversion from the Schur characters of the time variables $\{\bar p\}$ to those of the time variables $\{\bar p^\vee\}$ is done with a series of re-expansions using the symmetric group characters.

As the first step, one makes the plethystic transform of the OV partition function in terms of the same times $\{\bar p\}$:
\be\label{OVe}
Z_{OV}^{\cal K} = \sum_R  H_{_R}^{\cal K}(A,q)\,\chi_{_R}\{\bar p\}
= {\rm Plexp}\left(\sum_R f_{_R}(q,A)\chi_{_R}\{\bar p\}\right)
\ee
which is manifestly done with the inverse formula \cite{LM}
\be\label{fR}
f_{_R}^{\cal K}(A,q)=\sum_{d,m=1}(-1)^{m-1}{\mu (d)\over md}\sum_{\Delta_1,\ldots\Delta_m} \widehat{Ad}_d\,\psi_{_R}\Big(\sum_{i=1}^m\Delta_i\Big)\cdot\sum_{R_1,\ldots,R_m} \prod_{j=1}^m{\psi_{_{R_j}}(\Delta_j)\over z_{_{\Delta_j}}}H_{_{R_j}}^{\cal K}(A^d,q^d)
\ee
where the sum of two Young diagrams $\Delta$ and $\Delta'$ is the Young diagram with the lines $\{\delta_i,\delta_i'\}$ with a proper reordering, $\widehat{Ad}_d\,\Delta=\widehat{Ad}_d\,\{\delta_i\}=\{d\delta_i\}$, and $\mu(d)$ is the M\"obius function defined as follows: if the prime decomposition of $d$ consists of $m$ multipliers and contains non-unit multiplicities, $\mu(d)=0$, otherwise $\mu(d)=(-1)^m$.

At the second step, one needs to re-expand the r.h.s. of (\ref{OVe}) in the characters of
time variables $\{\bar p^\vee\}$. To this end, one has to use the expansion of the Schur characters into monomials
\be
\chi_{_R}\{\bar p\}=\sum_{\Delta}{\psi_{_R}(\Delta)\over z_{_\Delta}}\bar p_\Delta
\ee
where $\bar p_{\Delta}=\prod_i \bar p_i^{\,\mu_i}$, and the orthogonality relations for the symmetric group characters
\be\label{orth}
\sum_R {1\over z_{_\Delta}}\psi_{_R}(\Delta)\psi_{{_R}}(\Delta ')=\delta_{\Delta\Delta '},\ \ \ \ \ \sum_\Delta {1\over z_{_\Delta}}\psi_{_R}(\Delta)\psi_{_{R'}}(\Delta)=\delta_{_{RR'}}
\ee
Then, one can re-expand
\be
\chi_{_R}\{\bar p\}=\sum_Q C_{_{RQ}}\chi_{_Q}\{\bar p^\vee\}
\ee
with
\be\label{CRQ}
C_{_{RQ}}=\sum_\Delta {\psi_{_R}(\Delta)\psi_{_Q}(\Delta)\over z_{_\Delta}}\prod_i \Big(q^{\delta_i}-q^{-\delta_i}\Big)^{-1}
\ee
and write at the r.h.s. of (\ref{OVe})
\be
\sum_R f_{_R}(q,A)\chi_{_R}\{\bar p\}=\sum_{R,Q} f_{_R}(q,A)C_{_{RQ}}\chi_{_Q}\{\bar p^\vee\}
\ee
In other words, in order to find the integers $N_{Q,g,n}^{\cal K}$ one has to expand, using (\ref{fR}) and (\ref{CRQ}) the combination
\be
\sum_R f_{_R}C_{_{RQ}}=\frac{1}{q-q^{-1}}
\sum_{Q,g,n}  N_{Q,g,n}^{\cal K}\cdot (q-q^{-1})^{2g}\, A^n
\ee

\subsection{Gaussian distribution}

The next point to remind is that the Gaussian distribution is actually
an avatar of the binomial one for large $N$: it describes the behavior
near the maximum.
It is important that {\it large} $N$ does not need to be too large: the
numbers of the order of tens are more than sufficient.

\paragraph{The binomial distribution}
\be
C_{2N}^{K} p^{K}(1-p)^{2N-K}=
{(2N)!\over (2N-K)!K!} p^{K}(1-p)^{2N-K}
\ee
is peaked at $K=2pN$,
where
\be
C_{2N}^{N+k} p^{N+k}(1-p)^{N-k} \approx {1\over 2\sqrt{\pi pqN}}\cdot
\exp\left({-\frac{(k-(2p-1)N)^2}{4pqN}}\right)
\ee
i.e. it looks like {\bf the Gaussian distribution}
\be
{e^{-\frac{(k-\mu)^2}{2\sigma^2}}\over\sqrt{2\pi\sigma}}
\ee
with
\be
\sigma = {\sqrt{2pqN}}, \ \ \ \ \
\mu = (2p-1)N
\ee

{\bf The Poisson distribution} arises from the binomial distribution
when $p\longrightarrow 0$, $N\longrightarrow \infty$,
$a=2pN$ fixed.
Then, non-vanishing remain only the contributions with $k\ll N$ and
\be
C_{2N}^K p^K(1-p)^{2N-K} \sim \frac{(2Np)^K}{K!}\left(1-\frac{2pN}{2N}\right)^{2N}
\sim \frac{a^Ke^{-a}}{K!}
\ee

\subsection{Gaussianity of LMOV numbers}

We can now return to the LMOV free energy
\be
{F}^{\cal K} = \frac{1}{q-q^{-1}}
\sum_{Q,g,n }
N^{\cal K}_{Q,g,n} A^n  C_Q\{\bar p\}\cdot (q-q^{-1})^{2g}
\label{fLMOV}
\ee
with
\be
C_Q\{\bar p\}=\frac{1}{q-q^{-1}}\sum_\Delta   {\psi_Q(\Delta)\over z_\Delta}\,
\prod_{i=1}^{l_\Delta} (q^{\delta_i}-q^{-\delta_i})\, \bar p_{\delta_i}
= \sum_\Delta   {\psi_Q(\Delta)\over z_\Delta}\,(q-q^{-1})^{l_\Delta-1}
\prod_{i=1}^{l_\Delta} [\delta_i]\, \bar p_{\delta_i}
\ee
and
$N_{Q,g,n}$ obeying the Gaussian distribution in $g$.
We now substitute them by the binomial distribution.
Of course, this is badly justified
because the relevant $N$ are
not too big and the discrete points are far from being dense
on the Gaussian curve.
However, a more favorable statement about (\ref{Gaussfit})
is that the LMOV numbers are not just Gaussian, they are {\it binomially} (!)
distributed:
\be
\boxed{
N^{\cal K}_{Q,g,n} \approx
B^{\cal K}_{Q,n}(g) \equiv
I^{\cal K}_{_{Q,n}}\cdot \frac{N!}{g!(N-g)!} \cdot p^g(1-p)^{N-g}
}
\label{binomfit}
\ee
with $I$, $N$ and $p$ depending on ${\cal K}$, $n$ and $Q$.
Note that $p=p^{\cal K}_{_{Q,n}}$ can slightly deviate  from $1/2$,
and $N=N^{\cal K}_{_{Q,n}}$, from the number of points
(from the power of the polynomial in $q$ of the $A^n$-contribution to the LMOV polynomial).

Formally, substituting this distribution into (\ref{fLMOV}), we obtain
\be
F \approx \frac{1}{z}
\sum_{Q,n} \tilde I_{Q,n} C_Q\{\bar p\}\, A^n \cdot (1+\eta_{_{Q,n}} z^2)^{N_{Q,n}}
= \frac{1}{z} \sum_{R,Q,n} C_{RQ}(q)\,A^n\,\chi_{_R}\{\bar p\}\cdot \tilde I_{_{Q,n}}\cdot
(1+\eta_{_{Q,n}} z^2)^{N_{Q,n}}
\ee
with $z=q-q^{-1}$, $\eta = \frac{p}{1-p}$ and $\tilde I = I\cdot (1-p)^N$.
We omit the knot label ${\cal K}$ from $I$ and $\eta$ to simplify the formula, at least a little.
Thus, at any $R$ and $n$, we describe the generating function for the Ooguri-Vafa numbers
as a sum of a few powers of $(1+\eta\,z^2)$: the number of terms is equal to the number
of diagrams $Q$ of the size $|R|$. This describes the plot in Fig.\ref{OVexa} as a sum of several Gaussians.
Analytically, this provides an approximate formula for the multiplicities $M_\omega$ in
(\ref{strefcount}) and their generating function as a sum of several powers of a
peculiar variable surprisingly close to
\be
(1+z^2) = q^2-1+q^{-2} = \frac{[6]}{[3][2]} = \frac{(q^2-e^{2\pi i/3})(q^2-e^{-2\pi i/3})}{q^2}
\ee
Unfortunately, it is not immediately clear how this observation can be rigorously formulated. In s.4, we treat the Gaussian/binomial distributions in a slightly more sophisticated way, which turns out to be surprisingly successful for interpretation of the experimental data surveyed in the next s.3.

\section{Experimental evidence
\label{results}}

In this section we provide a few more gaussian/binomial plots to better illustrate
the phenomenon demonstrated in Fig.\ref{mainexa}. We also discuss various connected problems, in particular, dependencies of the numbers $N_{Q,g,n}^{\cal K}$ not only on the genus $g$, but also on $n$ and on the knot ${\cal K}$, the latter for the natural series of 2-strand and 3-strand torus knots and twist knots and for different representations. We realize an interesting universal behaviour for the dependence on the series of knots.

\subsection{Gaussian $g$-dependence}

First of all, we illustrate the phenomenon of the Gaussian distribution of the LMOV numbers already demonstrated in Fig.\ref{mainexa} for knot $8_{20}$. Since the Gaussian and binomial curves are practically coincide in all the cases considered below, from now on, we cite as an example only the Gaussian curves, though indicating parameters of the binomial distribution in each case.

\begin{figure}[b]
\bigskip
\includegraphics[bb = 0 0 15cm 17cm,width=150pt,height=100pt]{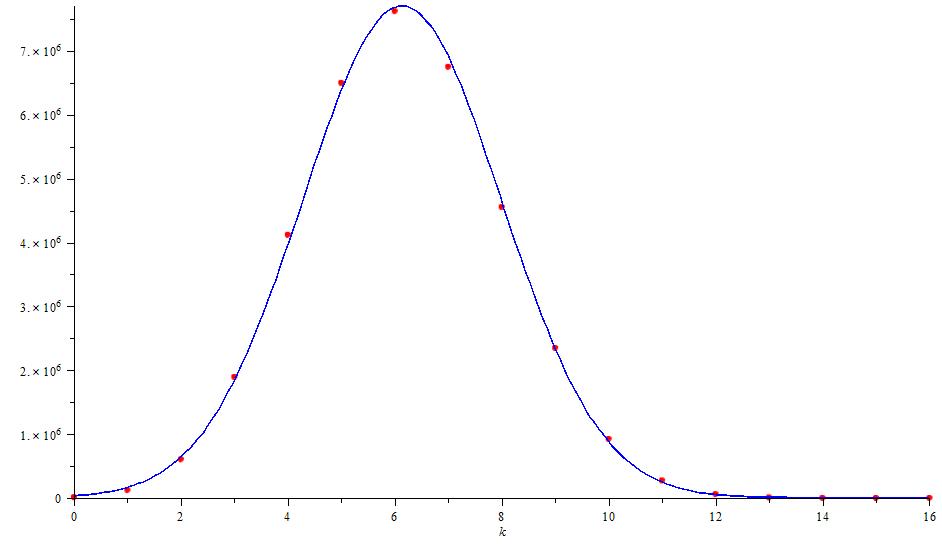}
\caption{\footnotesize
The LMOV numbers as a function of genus for the trefoil and the representation $Q=[4]$ and $n=16$: $N_{[4],g,16}^{3_1}$.
The blue curve is the Gaussian curve (\ref{Gaussfit}) with parameters $\mu=6.14$, $\sigma=1.85$, $I=3.58\cdot 10^7$, while the parameters of the binomial distribution are $p=0.44$, $n=14.0$.
}
\label{trefoil}
\end{figure}

Our examples start with the simplest knot, trefoil, Fig.\ref{trefoil}. It is a typical picture, the figures for other representations and other values of $n$ look much similar, with a light variation of parameters (see the next subsection).

\begin{figure}[t]
\bigskip
\includegraphics[bb = 0 0 15cm 17cm,width=150pt,height=100pt]{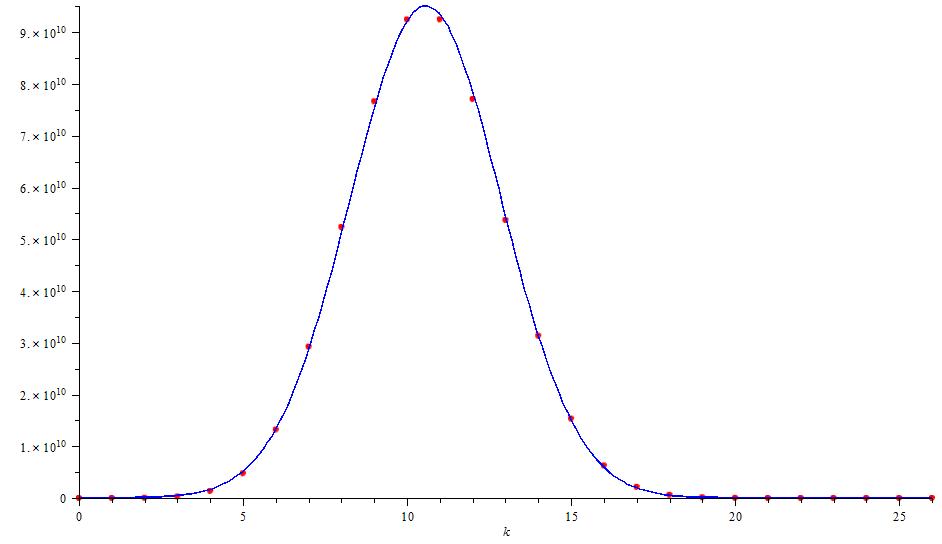}
\caption{\footnotesize
The LMOV numbers as a function of genus for knot $8_{20}$ and the representation $Q=[4]$ and $n=-16$: $N_{[4],g,-16}^{8_{20}}$.
The blue curve is the Gaussian curve (\ref{Gaussfit}) with parameters $\mu=10.57$, $\sigma=2.31$, $I=5.50\cdot 10^{11}$, while the parameters of the binomial distribution are $p=0.50$, $n=21.3$.
}
\label{820}
\end{figure}

In order to illustrate that the concrete value of $n$ does not matter too much, we also give a distribution of the LMOV numbers at $n=-16$, Fig.\ref{820}, which should be compared with Fig.\ref{mainexa} corresponding to $n=0$.

\begin{figure}[t]
\bigskip
\includegraphics[bb = 0 0 25cm 17cm,width=200pt,height=100pt]{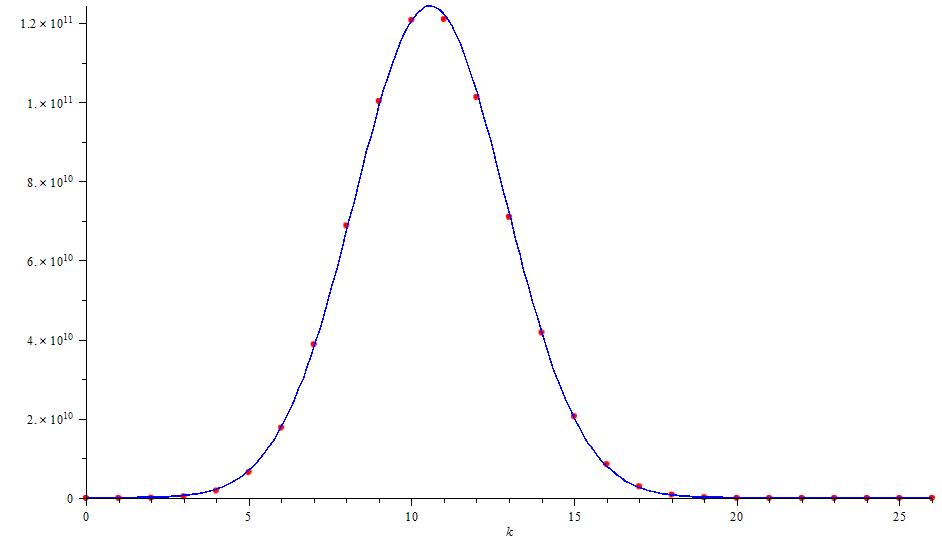}
\includegraphics[bb = 0 0 25cm 17cm,width=200pt,height=100pt]{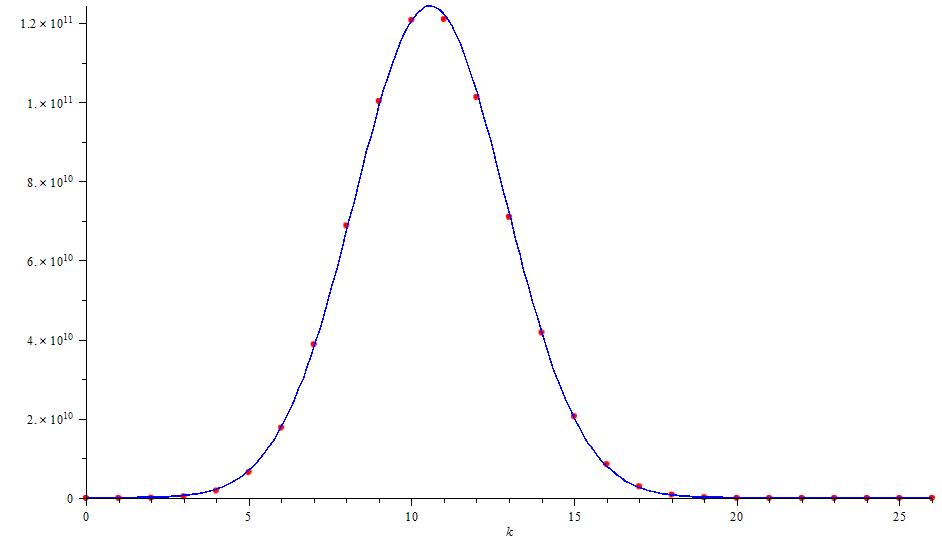}
\caption{\footnotesize
The LMOV numbers as a function of genus for the mutant pair of knots $11n41$ (left) and $11n47$ (right) and the representation $Q=[3]$ and $n=-15$: $N_{[3],g,-15}$. The blue curves are the Gaussian curves (\ref{Gaussfit}) with parameters $\mu=11.34$, $\sigma=2.40$, $I=6.85\cdot 10^{12}$, while the parameters of the binomial distributions are $p=0.494$, $n=22.98$ and $p=0.506$, $n=22.64$ correspondingly.
}
\label{11n41}
\end{figure}

As our third example, we demonstrate the LMOV number distributions for more complicated knots: for the pair of mutants $11n41$ and $11n47$ (Fig.\ref{11n41}). We deal with representations at the third level, where the HOMFLY polynomials of this pair first become to differ (for the representation $R=[2,1]$ only). The difference of the LMOV numbers for these knots is also Gaussian distributed, which we discuss in s.5.

\subsection{$n$-dependence\label{nd}}

\begin{figure}[h]
\bigskip
\includegraphics[bb = 0 0 15cm 17cm,width=150pt,height=100pt]{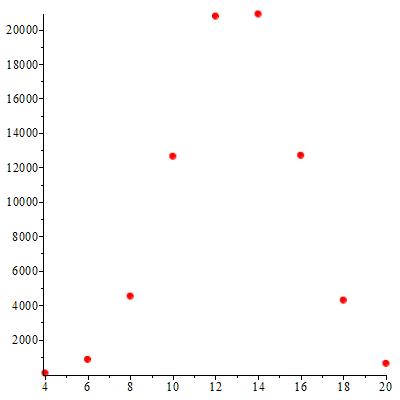}
\includegraphics[bb = 0 0 15cm 17cm,width=150pt,height=100pt]{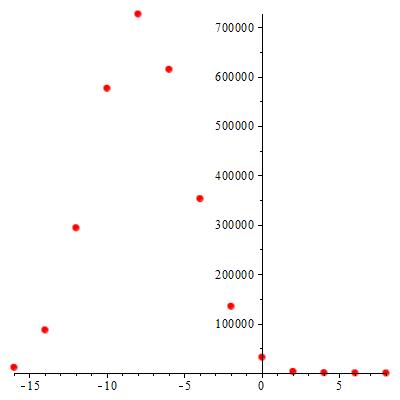}
\caption{\footnotesize
The LMOV numbers for knots $3_1$ (left) and $8_{20}$ (right) and the representation $Q=[4]$ as functions of $n$.
}
\label{31820}
\end{figure}

\begin{figure}[h]
\bigskip
\includegraphics[bb = 0 0 15cm 17cm,width=150pt,height=100pt]{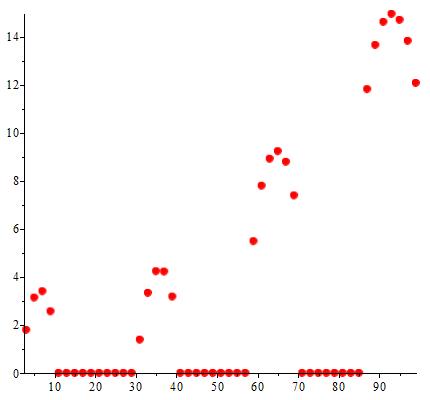}
\caption{\footnotesize
Distribution of logarithm of the LMOV numbers $\log\Big(N_{[3],0,n}^{Tw_{_{15}}}+1\Big)$ for the twist knot $Tw_{_{15}}$ and the representation $Q=[3]$ as a function of $n$ (for the sake of visualization).
}
\label{twn}
\end{figure}

Instead of the genus-related variable $g$ one can look at the charge-related $n$
and at the distribution of LMOV numbers $N_{Q,g,n}^{\cal K}$ in $n$ at fixed genus. We choose as examples the same knots $3_1$ and $8_{20}$ and minimal possible degree of $z$, since this is the case when the number of points is maximal. Unfortunately, this number is still quite little, nevertheless, one can see that the distribution is looking like a bell (Figs.\ref{31820}). This bell-like distribution typically happens for other knots too, with a notable exception: this is not the case for the twist knots (or just antiparallel braid, \cite{evo1}) at first representations and the number of twists large enough (see Fig.\ref{twn} for $Tw_{_{15}}$ and representation $[3]$).

\begin{figure}[h]
\bigskip
\includegraphics[bb = 0 0 15cm 17cm,width=150pt,height=100pt]{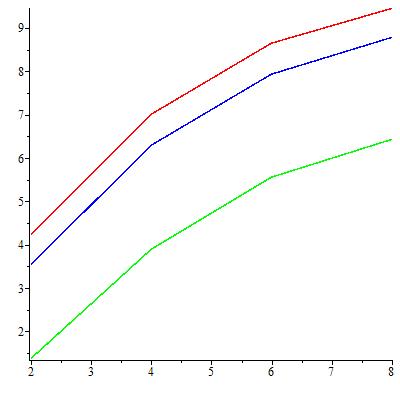}
\includegraphics[bb = 0 0 15cm 17cm,width=150pt,height=100pt]{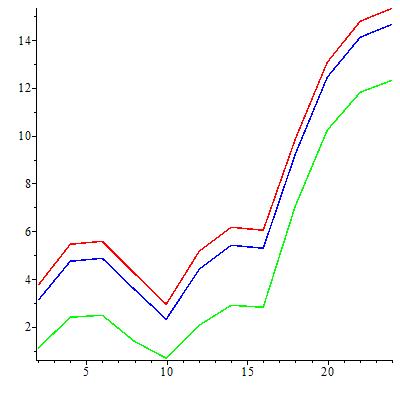}
\includegraphics[bb = 0 0 15cm 17cm,width=150pt,height=100pt]{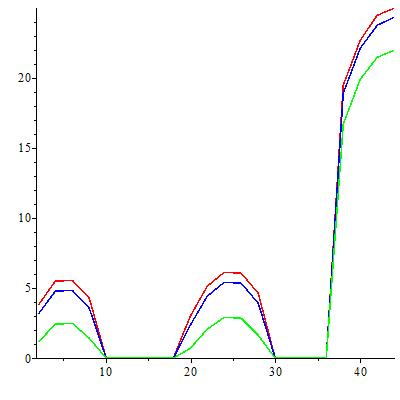}
\includegraphics[bb = 0 0 15cm 17cm,width=150pt,height=100pt]{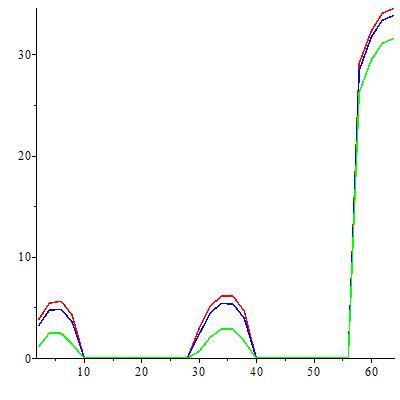}
\caption{\footnotesize
Plots of $\log(I_Q+1)$ as functions of $n$ for the twist knots $Tw_k$ at $k=1$ (first row, left), $k=5$ (first row, center), $k=10$ (first row, right) and $k=15$ (second row). The curves correspond to representations: red, to $Q=[3]$, blue, to $Q=[2,1]$, green, to $Q=[1,1,1]$.
}
\label{twist}
\end{figure}

Indeed, the family of twist knots can be treated by the evolution method \cite{DMMSS,evo1,evo2,evo3,MMf,Rama2}. In this case, the evolution goes along the antiparallel braid \cite{evo1} and the HOMFLY polynomial is generically given by the formula
\be
H_R^{Tw_k}(A,q) = \sum_{Q\in R\times \bar R} \Big(A^{\xi_Q}q^{\zeta_Q}\Big)^k h_Q(A,q)
\label{evoexpan}
\ee
with some definite $\xi_Q$ and $\zeta_Q$ depending only on the representation $Q$. This formula means that, at fixed $|R|$ and, hence, $|Q|=2|R|$ one can choose $k$ large enough in order to separate powers of $A$ and this separation is inherited by the plethystic free energy and by its LMOV counterpart.
In other words, for large enough $k$ exceeding the $A$-powers of particular $h_Q(A,q)$,
one can not have a single bell but several ones peaked around the mean-values $n_Q\sim l_Q\cdot k$. One can see this phenomenon of sliding apart with increasing $k$ at the plots $\log(I_Q,n+1)$ for $Tw_k$ as functions of $n$ at $k=1$, $5$, $10$, $15$ (Fig.\ref{twist}).

This argument is, in fact, extended to other families of knots containing the antiparallel braid.
However, if we now fix $k$ and let $|R|$ grow, the separate peaks begin to overlap
(because the $A$-powers of $h_Q$ grow with $Q$), and finally form a single-bell
distribution.

\subsection{Deviations from Gaussianity at small representations $Q$: $g$-distributions\label{3.3}}

\begin{figure}[h]
\bigskip
\includegraphics[bb = 0 0 15cm 17cm,width=150pt,height=100pt]{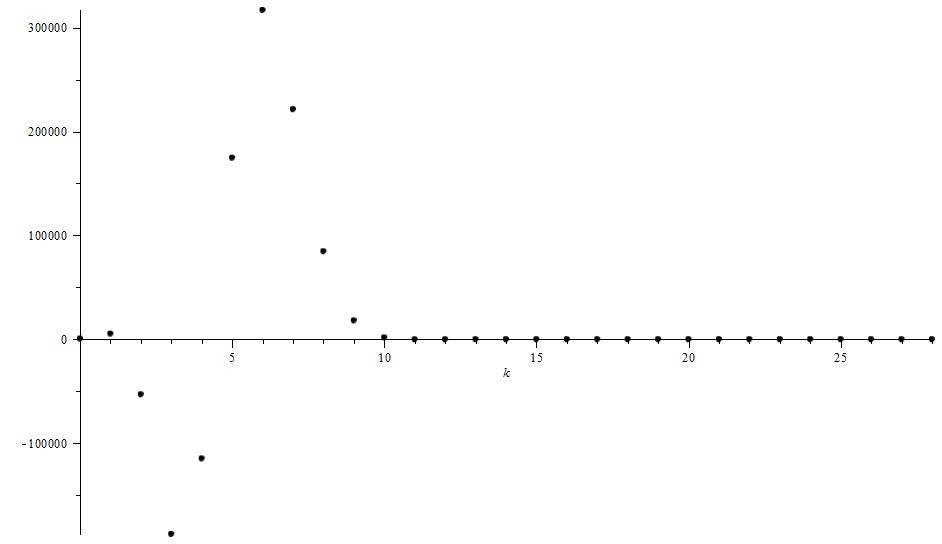}
\caption{\footnotesize
The antiparallel pretzel knot with 5 fingers: $(\bar 3,\bar 3,\bar 3,-\bar 3,-\bar 3)$ which forms a mutant pair with the pretzel knot $(\bar 3,\bar 3,-\bar 3,\bar 3,-\bar 3)$ \cite{MMMRS}. The LMOV numbers are given here as a function of genus for $Q=[2,1]$ and $A^{-1}$: $N_{[2,1],g,-1}$.
}
\label{pret}
\end{figure}

Moreover, the same argument as in the previous subsection, actually applies to distributions in $g$.
This is because the evolution depends not only on $A$ but also on $q$ as can be seen
in (\ref{evoexpan}), and, hence, $k$ separates also the peaks in $g$.
Strictly speaking, this time the switch from $q$ to $z$ obscures
the splitting with increasing $k$, however, the effect survives in some form (in particular, the sign at different genera becomes to change, see Fig.\ref{pret}) for some specific knots: for the pretzel knots with odd number of fingers and antiparallel braids \cite{evo3,MMMRS}. The knots of this kind also form mutant pairs \cite{MMMRS} that should be presumably distinguished only by the HOMFLY polynomials in representation $[4,2]$ \cite{Morton}, which is not accessible yet despite the recent progress \cite{MMM,Hcalc,arbor,evo3,MMMRS,MMf,Rama2}. Hence, we illustrate the phenomenon only at the level 3 of representations, Fig.\ref{pret}. Nevertheless, we expect that the same phenomenon as above will take place: with increase of the representation
the distribution becomes more and more Gaussian.

\bigskip

In other words, our observation of Gaussianity requires the limit of
large representations:
\be
  |R|\gg \ \text{regularity (typical evolution parameters) of substructures
inside the knot}
\ee
For well-structured knots, and even for those which have some pronounced
sub-structure like a long braid, parallel or antiparallel, one can need to go
to high enough $|R|$ to observe Gaussianity, it is not reached homogeneously
for all knots at once.

\subsection{$\mu$, $\sigma$ and $I$ as functions of $n$}

\begin{figure}[h]
\bigskip
\includegraphics[width=140pt,height=150pt]{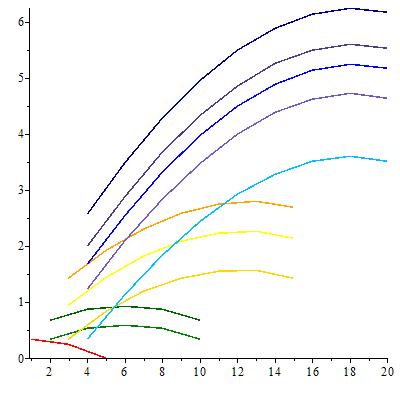}
\includegraphics[width=140pt,height=150pt]{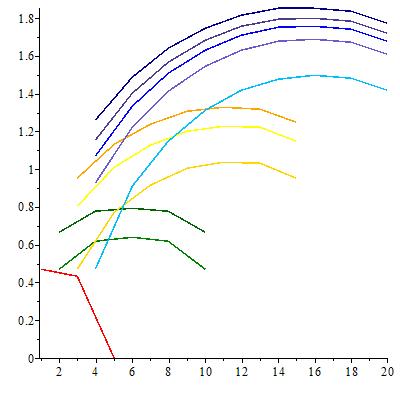}
\includegraphics[width=140pt,height=150pt]{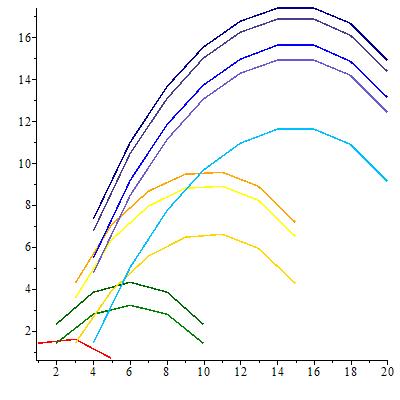}
\caption{\footnotesize
The plots of parameters of the Gaussian curve describing the LMOV numbers against $n$ for the trefoil. The left plot describes the dependence of the average $\mu$, the central plot, of the dispersion $\sigma$ and the right plot, of the integral $I$. Different levels of representations are drawn by different colors: red is for the first level (one representation $[1]$), green, for the second level (representations correspondingly $[1,1]$ and $[2]$ upward), yellow, for the third level (representations correspondingly $[1,1,1]$, $[2,1]$ and $[3]$ upward) and blue, for the fourth level (representations correspondingly $[1,1,1,1]$, $[2,1,1]$, $[2,2]$, $[3,1]$ and $[4]$ upward).
}
\label{31n}
\end{figure}

\begin{figure}[h]
\bigskip
\includegraphics[width=140pt,height=150pt]{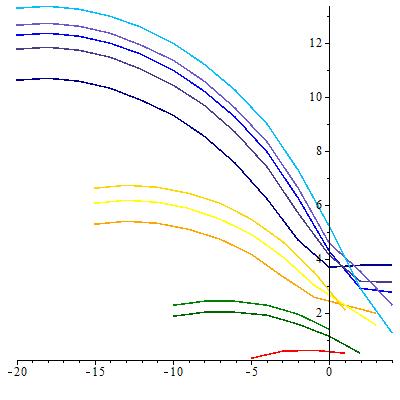}
\includegraphics[width=140pt,height=150pt]{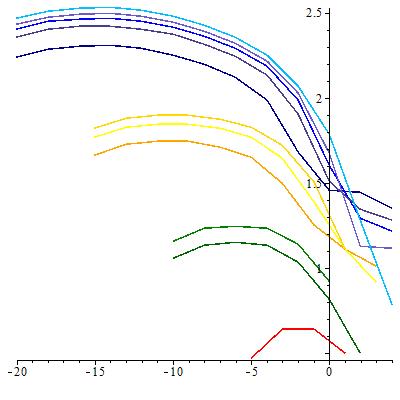}
\includegraphics[width=140pt,height=150pt]{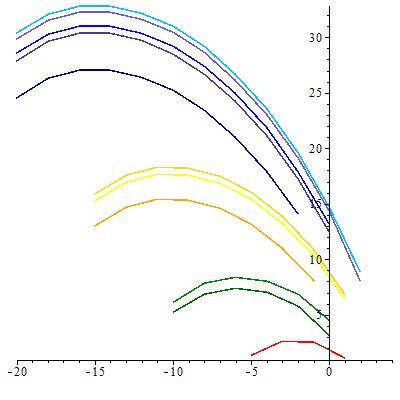}
\caption{\footnotesize
The plots of parameters of the Gaussian curve describing the LMOV numbers against $n$ for knot $8_{20}$. The left plot describes the dependence of the average $\mu$, the central plot, of the dispersion $\sigma$ and the right plot, of the integral $I$. Different levels of representations are drawn by different colors: red is for the first level (one representation $[1]$), green, for the second level (representations correspondingly $[1,1]$ and $[2]$ upward), yellow, for the third level (representations correspondingly $[1,1,1]$, $[2,1]$ and $[3]$ upward) and blue, for the fourth level (representations correspondingly $[1,1,1,1]$, $[2,1,1]$, $[2,2]$, $[3,1]$ and $[4]$ upward).
}
\label{820n}
\end{figure}

In this subsection, we illustrate how the parameters of the Gaussian distribution $\mu$, $\sigma$ and $I$ depend on $n$. The plots of dependence on $n$ are drawn in Figs.\ref{31n}-\ref{820n}. We again cite as examples the trefoil and knot $8_{20}$.

One can see again that, without special reasons that we discussed in s.\ref{nd}, the distributions are like a single bell. One can compare these pictures with Fig.\ref{twist} of s.\ref{nd}.

\subsection{$\mu$ and $\sigma$ as functions of knots\label{3.5}}

In this subsection, we illustrate how the parameters of the Gaussian distribution $\mu$ and $\sigma$depend on the type of knot. To make this question sensible, more than just a chaotic list of values for particular knots one needs a clever way to enumerate knots. This is a biggest classification problem, unsolved in knot theory for decades. An approach we find most fruitful is based on consideration of the evolution families \cite{DMMSS,evo1,evo2,evo3,MMf,Rama2}, when the knot diagrams are made from elementary constituents like various sub-braids and "fingers"  with their own "evolution parameters". Dependence of knot polynomials on these parameters can be calculated in an efficient way and this provides very clear control over the knot calculus.  The only drawback of this approach is overcounting: a particular knot/link can be a member of different families (families overlap), but for many purposes (except for the rigorous classification theorems) this does not cause a problem. Embedding of more traditional lists like Rolfsen table and its extensions \cite{katlas} into several simple families is described in \cite{Rama2}. In this paper, we consider only evolutions in particular directions, this is enough for our purposes here. However, generalizations are straightforward.

The most natural series of knot is given by a set of knots with a braid of changing length (this kind of series of knots has been studied in \cite{evo1,evo2,evo3,MMf,Rama2} by the evolution method). The simplest such series are torus knots (in the case of the parallel braid) and the twist knots (in the case of antiparallel braid), more complicated though still natural are double braid series \cite{evo1} and pretzel knots and links \cite{evo3}. Here we consider only the first two examples: the 2- and 3-strand torus knots, i.e. the torus knots $T[2,2k+1]$ and $T[3,3k\pm 1]$, and the twist knots $Tw_k$.

We start with the study of $T[2,2k+1]$-series of the torus knots. It appears that the parameters of the Gaussian curve as a function of $k$ for the genus chosen in such a way that $I_{Q,n}$ is maximal linearly depend on $k$ with a very high accuracy. The corresponding value of $n$ turns out to be given by a very simple linear function (note that one can instead choose $n$ in such a way that $n-n_{min}$ is fixed, where $n_{min}$ is the minimal $n$ when the LMOV coefficients are non-zero; the linear behaviour with such a choice still persists). For the representations at the third level, the values of $n$ when $I_{Q,n}$ is maximal are $6k+2$, and
the linear dependence is as follows
\begin{itemize}
\item For representation $Q=[3]$
\begin{equation}
\begin{array}{l}
\mu_{[3],6k+2}=4.02k-1.47
\\
\sigma^2_{[3],6k+2}=1.62k+0.36
\end{array}\nn
\end{equation}
\item
For representation $Q=[2,1]$
\begin{equation}
\begin{array}{l}
\mu_{[2,1],6k+2}=4.02k-2.03
\\
\sigma^2_{[2,1],6k+2}=1.62k+0.16
\end{array}\nn
\end{equation}
\item
For representation $Q=[1,1,1]$
\begin{equation}
\begin{array}{l}
\mu_{[1,1,1],6k+2}=4.02k-2.78
\\
\sigma^2_{[1,1,1],6k+2}=1.62k-0.22
\end{array}\nn
\end{equation}
\end{itemize}

Similarly, for the representations at the fourth level the straight lines are:
\begin{itemize}
\item
For representation $Q=[4]$
\begin{equation}
\begin{array}{l}
\mu_{[4],8k+3}=7.15k-1.27
\\
\sigma^2_{[4],8k+3}=2.87k+0.74
\end{array}\nn
\end{equation}
\item
For representation $Q=[3,1]$
\begin{equation}
\begin{array}{l}
\mu_{[3,1],8k+3}=7.15k-2.22
\\
\sigma^2_{[3,1],8k+3}=2.87k+0.42
\end{array}\nn
\end{equation}
\item
For representation $Q=[2,2]$
\begin{equation}
\begin{array}{l}
\mu_{[2,2],8k+3}=7.15k-1.85
\\
\sigma^2_{[2,2],8k+3}=2.87k+0.57
\end{array}\nn
\end{equation}
\item
For representation $Q=[2,1,1]$
\begin{equation}
\begin{array}{l}
\mu_{[2,1,1],8k+3}=7.15k-2.74
\\
\sigma^2_{[2,1,1],8k+3}=2.87k+0.20
\end{array}\nn
\end{equation}
\item
For representation $Q=[1,1,1,1]$
\begin{equation}
\begin{array}{l}
\mu_{[1,1,1,1],8k+3}=7.15k-3.88
\\
\sigma^2_{[1,1,1,1],8k+3}=2.87k+0.36
\end{array}\nn
\end{equation}
\end{itemize}

\begin{figure}[h]
\bigskip
\includegraphics[width=140pt,height=150pt]{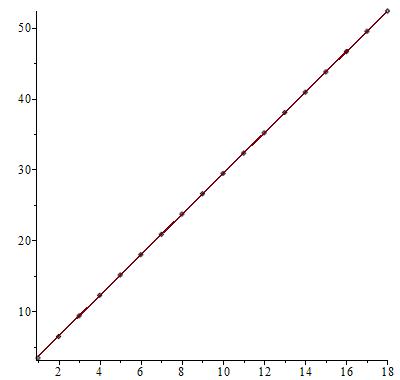}
\hspace{3cm}\includegraphics[width=140pt,height=150pt]{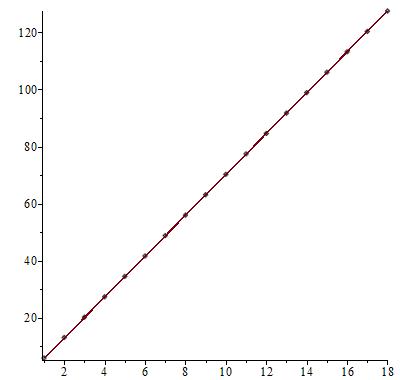}
\caption{\footnotesize
The plots of parameters $\mu$ (the left figure) and $\sigma^2$ (the right figure) for the series of torus knots $T[2,2k+1]$ as a function of $k$. Plotted are
$\mu_{[4],8k+3}$ and $\sigma^2_{[4],8k+3}$.
}
\label{2slin}
\end{figure}

One can see that the slopes of the straight lines depend only on the level of the representation, in variance with the constant shifts. For illustrative purposes, we demonstrate the corresponding plots in Fig.\ref{2slin} for $\mu_{[4],8k+3}$ and $\sigma^2_{[4],8k+3}$.

Similarly, linear is the dependence on the knot for the 3-strand torus knots: for the series $T[3,3k+1]$ the linear dependence is

\begin{equation}
\begin{array}{lcl}
\mu_{[3],18k+9}=12.07 k-1.89,
&\ &
\sigma_{[3],18k+9}=4.84 k+0.59
\\
\mu_{[2,1],18k+9}=12.07 k-2.46,
&\ &
\sigma_{[2,1],18k+9}=4.84 k+0.40
\\
\mu_{[1,1,1],18k+9}=12.07 k-3.23,
&\ &
\sigma_{[1,1,1],18k+9}=4.84 k+0.05
\\
\mu_{[4],24k+12}=21.46 k-1.36,
&\ &
\sigma_{[4],24k+12}=8.60 k+1.17
\\
\mu_{[3,1],24k+12}=21.46 k-2.01,
&\ &
\sigma_{[3,1],24k+12}=8.60 k+1.00
\\
\mu_{[2,2],24k+12}=21.46 k-2.37,
&\ &
\sigma_{[2,2],24k+12}=8.60 k+0.85
\\
\mu_{[2,1,1],24k+12}=21.46 k-2.90,
&\ &
\sigma_{[2,1,1],24k+12}=8.60 k+0.64
\\
\mu_{[1,1,1,1],24k+12}=21.46 k-4.05,
&\ &
\sigma_{[1,1,1,1],24k+12}=8.60 k+0.09
\end{array}
\end{equation}
One can again see that the slopes of the straight lines do not depend on the representations (only on its level), and only the shifts do. Moreover, {\bf the ratios of slopes at the levels 3 and 4, both for the average and for the dispersion, are equal to the same figure 0.56 for the both series of knots.}

Thus, one can parametrize the average and the dispersion for the representations of level $p$ as
\be
\mu^{(m)}_p=\alpha^{(m)}_p\cdot k+\gamma_{[Q],m,p},\ \ \ \ \ \ \Big(\sigma^{(m)}_p\Big)^2=\beta^{(m)}_p\cdot k+\gamma_{[Q],m,p}
\ee
In fact, this kind of behaviour is looking quite universal: we can consider other series of the knots with a growing braid, and the result will be the same. Let us consider two more examples: the example of the twist knots, where the braid is antiparallel \cite{evo1}, and the 3-strand braids of the form $\{\underbrace{1,1,1,\ldots,1,1}_{2k-1},2,1,2,1\}$ and $\{\underbrace{1,1,1,\ldots,1,1}_{2k-1},-2,-1,-2,-1\}$, where the braids are parallel \cite{evo1}.
In the first case, the parameters are:
\be
\alpha^{tw}_2=0.89,\ \ \ \ \ \beta^{tw}_2=0.36,\ \ \ \ \ \ \alpha^{tw}_3=2.68,\ \ \ \ \ \beta^{tw}_3=1.08
\ee
while, in the second case, they coincide with the 2-strand torus case. Hence, we come to the conclusion that the parameters $\alpha^{(m)}_p$, $\beta^{(m)}_p$ depend only on the type of the increasing braid and conjecture their dependence of the form
\be\label{p1}
\alpha^{(m)}_p\sim m(m-1)p^2,\ \ \ \ \ \beta^{(m)}_p\sim m(m-1)p^2
\ee
for the parallel braid and
\be
\alpha^{tw}_p\sim p(p-1),\ \ \ \ \ \beta^{tw}_p\sim p(p-1)
\ee
for the antiparallel one. In order to check (\ref{p1}), we checked it for the 3-strand torus knots series $T[3,3k+1]$ at the level 5 and realize that $\alpha^{(3)}_5=33.53$ and $\beta^{(3)}_5=13.43$, in full accordance with (\ref{p1}). In s..\ref{evo} we give some theoretical arguments in favor of these formulas.

\section{Abundance and origins of Gaussian curves}

A natural question is how frequent are Gaussian distributions of coefficients,
is it really an exotics among "realistic" large polynomials with positive coefficients?
A suspicion can be that the Gaussian curve can be a good approximation for nearly any
function near its maximum, and thus can seem to be a very general phenomenon.
However, we are interested in the {\it coefficients}, not the function itself.
In other words, what should have a sharp maximum, is a kind of a Fourier transform
of the original function.
Thus, what one can expect from the coefficients?

\subsection{The role of polynomiality}

On one hand, if the function is of a general type, it is bounded and has
bounded derivatives, a typical example is something like $\sin x$.
Then, by the Maclaurin formula, the coefficients are derivatives divided by factorials,
i.e. typically they are  {\it small} rather than large.

On another hand, if we, say, take a monomial $x^M$, i.e. a function just with a single
non-vanishing coefficient, and then merely shift the argument, $x \longrightarrow x+1$,
the new function $(x+1)^M$ has binomially distributed coefficients
with a sharp maximum at large $M$, and this also looks quite typical.
The difference with the previous argument is that the $k$-the derivative
is $\sim M^k x^{M-k}$ and is in fact very big for any non-vanishing $x$
(for $x\gg M^{-1}$, to be precise).

A possible resolution of the seeming contradiction between these two expectations is that
the peaked coefficient distribution is a property of {\it polynomials},
for random polynomial of high degree the coefficients are typically large:
even if they were not such for some particular ("initial") choice of $x$,
they will get such after any shift of $x$.

If this new expectation is true, it looks like the question for polynomials
should probably be inverted: how at all can it happen that a polynomial
has high degree and moderate coefficients?
At the same time, {\it if} for {\it some} choice of $x$ the coefficients were
moderate, for the shifted $x$ they get binomially peaked and typically
there will be one dominating peak, simply because different binomial terms
have exponentially different heights, and just one will easily exceed
all the others.

Thus, we are led to a conclusion that the Gaussianity of the LMOV numbers
can be a {\it natural} corollary of just {\it the polynomiality} of the
original HOMFLY invariant and so is $f_R$, and miraculous is just the opposite:
that there is a peculiar variable $q$ in which they are polynomials
with moderate coefficients.
This last fact is easily understood for "regular" (well structured) knots,
where polynomials in $q$ get high degree as a result of evolution along
various braid lengths which basically contribute just a high power of $q$
without any non-trivial coefficient.
To search for HOMFLY and $f_R$ with large coefficients one should look
at very knotted (unstructured) knots which are realized as a closed braid with roughly the same
large number of strands as the number of intersections, and not regular
(like twist knots).

Still, even in this case one expects that increase in {\it the representation}
$R$ would restore "the regularity" and simplicity of the HOMFLY polynomial in the variable $q$,
and, thus, the Gaussianity of coefficient distributions in the variable $z$
occur at large enough $R$.

We will now provide some more technical illustrations to above arguments
with the very simple example: quantum numbers, which, in a certain sense, can be
considered as good models of characters in general and particularly of knot polynomials
(which are also characters, though of huge loop algebras).
This kind of language can also be useful to study the {\it deviations}
from Gaussianity, which, according to our general reasoning should also
be Gaussian: after subtraction of the main peak the sub-leading should
start dominating and so on.
We illustrate this by an example of {\it difference} between mutant knots,
which is conceptually related to correction issue, while technically,
to the quantum numbers.

\subsection{A toy example: quantum numbers}

First of all, a quantum number is a sharply-peaked function on the unit circle
\be
[M] \equiv \frac{q^M-q^{-M}}{q-q^{-1}} \ \stackrel{q=e^{i\phi}}{=}\
\frac{\sin(M\phi)}{\sin\phi}
\ee
This is because $[M]$ is a sum $\sum_{i=-M+1}^M q^{2i-1}$ with unit coefficients,
i.e. almost a Fourier transform of unity, which would be an infinitely-peaked
$\delta$-function.
Thus, it is clear that, in the variable $\phi$, we should get a nearly Gaussian
distribution.
But what does it have to do with $(1+z^2)^M = \left(1-4\sin^2\phi\right)^M$?

The quantum number $[M]$ can be rewritten\footnote{The simplest way to prove it is to use
\be
[M]=[2K+1]={\sin (2K+1)\phi\over\sin\phi}={\Im (\cos\phi+i\sin\psi)^{2K+1}\over\sin\phi}=
\sum_{l=0}^K\left(\begin{array}{c} 2K+1\\ 2l+1\end{array}\right) (-1)^l\sin^{2l}\phi\cdot\cos^{2K-2l}\phi=\nn\\=\sum_{l=0}^K\left(\begin{array}{c} 2K+1\\ 2l+1\end{array}\right) z^{2l}(1+z^2/4)^{k-l}
=\sum_{k=0}^K\sum_{l=0}^k\left(\begin{array}{c} 2K+1\\ 2l+1\end{array}\right) \left(\begin{array}{c} K-l\\ k-l\end{array}\right)\Big(z^2/4\Big)^{k}\nn
\ee
and to use the identity
$$
\sum_{l=0}^k\left(\begin{array}{c} 2K+1\\ 2l+1\end{array}\right) \left(\begin{array}{c} K-l\\ k-l\end{array}\right)={4^k(2K+1)\over (2k+1)!}{(K+k)!\over (K-k)!}
$$
} as a polynomial in $z$ for any odd $M=2K+1$:
\be\label{qnd}
[M]=[2K+1]= (2K+1)\sum_{k=0}^\infty \frac{z^{2k}}{(2k+1)!}\cdot \frac{(K+k)!}{(K-k)!}=\sum_{k=0}^\infty n_kz^{2k}
\ee
with $z = q-q^{-1}$. The coefficient $n_k$ in front of $z^{2k}$ is a deformed binomial distribution with the average
\be
\mu={1\over 2}\left[{M\over\sqrt{5}}\cdot {q_*^M+q_*^{-M}\over q_*^M-q_*^{-M}}-1\right],\ \ \ \ \ q_*={\sqrt{5}+1\over 2},\ \ \ \ \ z_*=1
\ee
It has the Gaussian form at large $K$
\be\label{Gl1}
\boxed{n_k=\frac{2K+1}{(2k+1)!}\cdot \frac{(K+k)!}{(K-k)!}\sim I_K\exp\left[-{5\sqrt{5}\over 4K}\left(k
-{K\over\sqrt{5}}\right)^2\right]}
\ee
with
\be\label{Gl2}
\sigma=\sqrt{2K\over 5\sqrt{5}},\ \ \ \ \ \ \
\mu={K\over\sqrt{5}},\ \ \ \ \ \ I_K={1\over\sqrt{K}}\left({\sqrt{5}+1\over \sqrt{5}-1}\right)^K
\ee

\subsection{More examples: ratios of quantum numbers}

One can observe a similar behaviour not only for the quantum numbers, but also for their ratios that are still polynomials in $q$ and can be expanded into series in $z$. Consider two examples.

\paragraph{Numbers of the form $\displaystyle{[4K]\over [2K]}$.}

Similarly to the previous subsection, one can perform these ratios of quantum numbers as sums of the deformed binomial distribution of the form
\be\label{qnd2}
{[4K]\over [2K]}=2N\cdot\sum_{k=0}^K{(2K-k-1)!\over (2K-2k)!k!}\cdot z^{2K-2k} =
2K\cdot\sum_{k=0}^K C_{K+k}^{2k}\cdot {z^{2k}\over K+k}=\sum_{k=0}^K{z^{2k}\over (2k)!}\cdot{(K+k)!\over (K-k)!}\cdot{2K\over K+k}
\ee
This distribution is much similar to (\ref{qnd}) and, in the limit of large $K$, has the average $\mu=K/\sqrt{5}$ and approaches to the Gaussian distribution (\ref{Gl1}) with the same parameters (\ref{Gl2}).

\paragraph{Numbers of the form $\displaystyle{[2(2K+1)]\over [2]\cdot [2K+1]}$.}

Yet another example is given by the ratios of the form $[2(2K+1)]/[2][2K+1]$, which also can be treated in the same way: these numbers are expanded into sums of the deformed binomial distribution of the form
\be\label{2K}
{[2(2K+1)]\over [2]\cdot [2K+1]}=\sum_{k=0}^K{(2K-k)!\over (2K-2k)!k!}\cdot z^{2K-2k} =\sum_{k=0}^K C_{K+k}^{2k}\cdot z^{2k}=\sum_{k=0}^K{z^{2k}\over (2k)!}\cdot{(K+k)!\over (K-k)!}
\ee
This distribution is also very close to (\ref{qnd}) and (\ref{qnd2}) thus, in the limit of large $K$, it also has the average $\mu=K/\sqrt{5}$ and approaches to the Gaussian distribution (\ref{Gl1}) with the same parameters (\ref{Gl2}).

\paragraph{Numbers of the form $\displaystyle{[2K]\over [2]}$.}

Similarly to the previous subsections, one can perform these ratios of quantum numbers as sums of the deformed binomial distribution of the form
\be\label{qnd3}
{[2K]\over [2]}=\sum_{k=0}^K{(K+k-1)!\over (K-k)!(2k-1)!}\cdot z^{2k}=\sum_{k=0}^K C_{K+k-1}^{2k-1}\cdot z^{2k}=\sum_{k=0}^K{z^{2k}\over (2k)!}\cdot{(K+k)!\over (K-k)!}\cdot{2k\over K+k}
\ee
This distribution is much similar to (\ref{qnd}) and, in the limit of large $K$, has the average $\mu=K/\sqrt{5}$ and approaches to the Gaussian distribution (\ref{Gl1}) with the same parameters (\ref{Gl2}).

\subsection{Gaussianity of mutant difference}

Amusing example directly follows from the previous one
and the result of \cite{MortonM,MMMRS} for mutant knots.
The {\it difference} between the two mutants is also
Gaussian distributed (!).
This is simply because this difference is made from
the ratios of quantum numbers, at least in the simplest case,
where it is already calculated:
\be\label{diffmut}
1.&\qquad H^{11a19}_{[2,1]} - H^{11a25}_{[2,1]} = A^{-7}\cdot f(A,q)\cdot\dfrac{[14]}{[2][7]}\cdot\mathfrak{n} &
2.\qquad H^{11a24}_{[2,1]} - H^{11a26}_{[2,1]} = A^{-1} \cdot f(A,q)\cdot\dfrac{[14]}{[2][7]}\cdot\mathfrak{n} \nn\\
3.&\qquad H^{11a44}_{[2,1]} - H^{11a47}_{[2,1]} = A \cdot f(A,q)\cdot\frac{[8]}{[2]}\cdot\mathfrak{n}  &
4.\qquad H^{11a57}_{[2,1]} - H^{11a231}_{[2,1]} = A^{-5}\cdot f(A,q)\cdot\frac{[8]}{[2]}\cdot\mathfrak{n}  \nn \\
5.&\qquad H^{11a251}_{[2,1]} - H^{11a253}_{[2,1]} = A^{-1} \cdot f(A,q)\cdot\dfrac{[14]}{[2][7]}\cdot\mathfrak{n}  &
6.\qquad H^{11a252}_{[2,1]} - H^{11a254}_{[2,1]} = A^{-5} \cdot f(A,q) \cdot\dfrac{[14]}{[2][7]}\cdot\mathfrak{n} \nn \\
7.&\qquad H^{11n34}_{[2,1]} - H^{11n42}_{[2,1]} = A^{3} \cdot f(A,q)\cdot\dfrac{[14]}{[2][7]}\cdot\mathfrak{n}  &
8.\qquad H^{11n35}_{[2,1]} - H^{11n43}_{[2,1]} = A^{19} \cdot f(A,q)\cdot\mathfrak{n}   \\
9.&\qquad H^{11n36}_{[2,1]} - H^{11n44}_{[2,1]}=A^{-9}\cdot f(A,q)\cdot\mathfrak{n}   &
10.\qquad H^{11n39}_{[2,1]} - H^{11n45}_{[2,1]} = A^{-3}\cdot f(A,q)\cdot\dfrac{[14]}{[2][7]}\cdot\mathfrak{n} \nn \\
11.&\qquad H^{11n40}_{[2,1]} - H^{11n46}_{[2,1]}=A^{13} \cdot f(A,q)\cdot\mathfrak{n}  &
12.\qquad H^{11n41}_{[2,1]} - H^{11n47}_{[2,1]} =A^{-15}\cdot f(A,q)\cdot\mathfrak{n}  \nn \\
13.&\qquad H^{11n71}_{[2,1]} - H^{11n75}_{[2,1]} = A^{13} \cdot f(A,q)\cdot\dfrac{[7][8]}{[14]}\cdot\mathfrak{n}   &
14.\qquad H^{11n73}_{[2,1]} - H^{11n74}_{[2,1]} = A^{-3}\cdot f(A,q)\cdot\dfrac{[8]}{[2]}\cdot\mathfrak{n}    \nn \\
15.&\qquad H^{11n76}_{[2,1]} - H^{11n78}_{[2,1]}= A^{-15} \cdot f(A,q)\cdot\dfrac{[7][8]}{[14]}\cdot\mathfrak{n}    &
16.\qquad H^{11n151}_{[2,1]} - H^{11n152}_{[2,1]} = A^{-9} \cdot f(A,q)\cdot\dfrac{[14]}{[2][7]}\cdot\mathfrak{n}\nn
\ee
where, for the sake of brevity, we introduced standard factors $\mathfrak{n}:=\dfrac{[3]^2[14]}{[2][7]}$ and $f(A,q):=(q-q^{-1})^{11}\cdot  D_{3}^2D_{2} D_0 D_{-2}D_{-3}^2$ with $D_n=(Aq^n-A^{-1}q^{-n})/(q-q^{-1})$.
These differences are all ratios of the quantum numbers of the types considered in the previous subsection, which is not surprising since the Young diagram $[2,1]$ does not change under transposition, i.e. these HOMFLY polynomials are invariant w.r.t. $q\to -1/q$ and are expanded into $z$.

\begin{figure}[h]
\bigskip
\includegraphics[bb = 0 0 15cm 17cm,width=150pt,height=100pt]{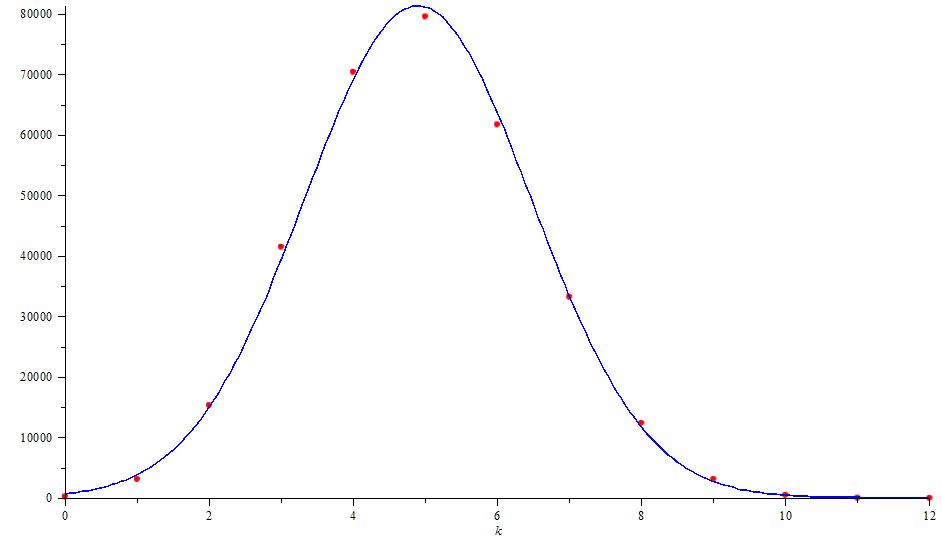}
\caption{\footnotesize
The differences of LMOV numbers
$N_{[2,1],g,-15}^{11n41}-N_{[2,1],g,-15}^{11n47}$ as a function of genus
for the mutant pair $11n41$ and $11n47$ (12th in the list (\ref{diffmut})).
The blue curve is the Gaussian curve (\ref{Gaussfit}) with parameters $\mu=4.90$, $\sigma=1.58$, $I=3.21\cdot 10^{5}$, while the parameters of the binomial distribution are $p=0.493$, $n=9.94$.
}
\label{mutadiff}
\end{figure}

Now we calculate the difference of the LMOV numbers for a pair of mutants at the third level. This difference is tiny at the level of each knot itself
(coefficients in the difference are much smaller than those in the full polynomials, see Fig.\ref{11n41}),
but it is well defined by itself and it is itself Gaussian,
see Fig.\ref{mutadiff}. In fact, since the HOMFLY polynomials of the mutant pair at the first three levels differs only for the representation $[2,1]$, the difference of the LMOV numbers can be read of directly from the difference (\ref{diffmut}), and the distribution of the numbers is read off from formulas of the previous subsection.

\subsection{Theoretical explanation and estimate: Gaussian parameters against knot\label{evo}}

Also, we can now provide a simple qualitative explanation for the dependence of the Gaussian average in the series of knots
empirically discovered and described in s.\ref{3.5}, which, despite being very rough, turns out to be surprisingly accurate.

Consider the torus knot $T[m,n]$. It can be realized by an $m$-strand closed braid with $nm$ crossings. The HOMFLY polynomial in the representation $R$ in this case is given by \cite{RJ}
\be\label{RJ}
H_R^{(T[m,n]}=\sum_{Q\in R^{\otimes m}} A^{-n(m-1)|R|}q^{mnC_2(R)-{n\over m}C_2(Q)} C_{RQ}D_Q
\ee
where are $C_{RQ}$ are numerical coefficients (describing the $m$-plethysm of the Schur functions), $D_Q(q,A=q^N)$ are the quantum dimensions of $SU(N)$ and $C_2(R)=\sum_i r_i(r_i-2l(R)+2i-1)$ is value of the second Casimir operator on $R$. The transition to the LMOV numbers mixes various representations $R$ of the given level, and the dominant contribution comes from the maximal value of the exponent of $q$ in (\ref{RJ}), which is given by the symmetric representation $R=[r]$ and the representation $Q=[\underbrace{r,r,\ldots,r}_{m}]$.
This contribution is proportional to $\displaystyle{q^{n(m-1)|R|^2}\over A^{n(m-1)|R|}}$ and, by the $q\to -1/q$ symmetry of the LMOV numbers, is restored to
the ratio of quantum numbers
\be
\sim {\left[{n(m-1)|R|^2}\right]_q\over [2]_q}
\ee
which can be expressed through the variable $z=q-q^{-1}$.
As we saw in s.4.3, the coefficients (\ref{qnd3}) of $z$-expansion of $[K]/[2]$
are well described by the Gaussian distribution with the average
$\mu = \frac{K}{2\sqrt{5}}$ and the dispersion $\sigma^2=2/5\cdot\mu$.

Now we can explain the results of s.\ref{3.5}. For instance, choosing $n=mk+i$, $i=1..m-1$, we obtain for the linear in $k$ parts of the average and the dispersion:
$\mu \sim \alpha\cdot k  $, $\sigma^2\sim\beta\cdot k$ with
\be
\alpha_{\!_{|R|}}^{(m)} \approx \frac{m(m-1)|R|^2}{2\sqrt{5}}
\label{alphaestim}
\ee
\be
\beta_{\!_{|R|}}^{(m)} \approx \frac{m(m-1)|R|^2}{5\sqrt{5}}
\approx \frac{2}{5}\alpha_{\!_{|R|}}^{(m)}
\label{betaestim}
\ee
and these $\alpha$, $\beta$ depend only in the size of representation $R$. We observed this phenomenon in s.\ref{3.5} for the torus knots $T[2,2k+1]$ and $T[3,3k+1]$.

This formula gives a very good estimates of $\alpha$'s and $\beta$'s, the concrete table of "predictions" (\ref{alphaestim}) and (\ref{betaestim}) is:
\be
\!\!\!\!\!\!\!\!\!\!\!\!\!\!\!\!\!\!\!\!\!\!\!\!\!
\alpha_{\!_{|R|}}^{(m)} =
\begin{array}{c|ccccccc}
m \backslash |R| & 1 & 2 & 3 & 4 & 5 & 6 & \ldots\\
\hline & \\
2 &0.45&1.79&4.02&7.16&11.2&16.1 \\
3 &1.34&5.37&12.07&21.47&33.54&48.30 \\
4 &2.68&10.73&24.15&42.93&67.08&96.60 \\
5 &4.47&17.89&40.25&71.55&111.8&161.0 \\
6 &6.71&26.83&60.37&107.3&167.7&241.5 \\
\ldots &
\end{array}
\beta_{\!_{|R|}}^{(m)} =
\begin{array}{c|ccccccc}
m \backslash |R| & 1 & 2 & 3 & 4 & 5 & 6 & \ldots\\
\hline & \\
2 &0.18&0.72&1.61&2.86&4.47&6.44 \\
3 &0.54&2.15&4.83&8.59&13.42&19.32 \\
4 &1.07&4.29&9.66&17.17&26.83&38.64 \\
5 &1.79&7.16&16.10&28.62&44.72&64.40 \\
6 &2.68&10.73&24.15&42.93&67.08&96.60 \\
\ldots &
\end{array}
\ee

These figures are completely supported by the experimental results from s.\ref{3.5}. Note also that from (\ref{alphaestim})-(\ref{betaestim}) it follows that
the ratios:
\be
\frac{\alpha^{(m)}_{\!_{|R|}}}{\alpha^{(m)}_{\!_{|R|+1}}} \approx \frac{\beta^{(m)}_{\!_{|R|}}}{\beta^{(m)}_{\!_{|R|+1}}}\approx \frac{|R|^2}{(|R|+1)^2}
\ee
do not depend on $m$, which was also observed in s.\ref{3.5}.

At last, we can notice that the $A$-dependence in (\ref{RJ}) is determined by the factor $A^{-n(m-1)|R|}$. This perfectly agrees with the values of $n$ when the $I_{Q,n}$ is maximal in s.\ref{3.5}: they are described by the linear dependence on the evolution parameter $k$, and the slope is given by $2|R|$ for the $T[2,2k+1]$ torus knots and by $6|R|$ for the $T[3,3k+1]$ torus knots.

Now note that this our torus knot consideration can be literally repeated for any parallel two-strand braid of the length $k$ at any knot (see \cite[Eq.(78)]{evo1}). Hence, the results observed in s.\ref{3.5} for the non-torus knot.

Another example considered in s.\ref{3.5} is the evolution along an antiparallel 2-strand braid as an example realized in the twist knots.
Then, the only change
in above formulas is in the $R$-dependence \cite{evo1}: in this case, the intermediate representation $Q\in R\otimes\bar R$ that gives the leading contribution is $Q=[2r,\underbrace{r,r,\ldots,r}_{N-2}]$ and the corresponding contribution is $\left(Aq^{r-1}\right)^{2kr}$ (see \cite[Eq.(112)]{evo1}), i.e. the answers are
\be
\alpha_{\!_{|R|}}^{tw} \approx \frac{ |R|\,\big(|R|-1\big)}{\sqrt{5}},
\ \ \ \ \ \
\beta_{\!_{|R|}}^{tw} \approx \frac{ 2|R|\,\big(|R|-1\big)}{10\sqrt{5}}
\ee
which gives
\be
\alpha_{\!_{2}}^{tw} = \frac{2}{\sqrt{5}} = 0.89, \ \ \ \
&\alpha_{\!_{3}}^{tw} = \frac{6}{\sqrt{5}} = 2.68, \ \ \ \
&\alpha_{\!_{4}}^{tw} = \frac{12}{\sqrt{5}} = 5.37 \nn \\
\beta_{\!_{2}}^{tw} = \frac{4}{5\sqrt{5}} = 0.36, \ \ \ \
&\beta_{\!_{3}}^{tw} = \frac{12}{5\sqrt{5}} = 1.07, \ \ \ \
&\beta_{\!_{4}}^{tw} = \frac{24}{5\sqrt{5}} = 2.15
\ee
in full accordance with the experimental results of s.\ref{3.5}.

Note that the $A$-dependent factors in this case depend also on the representation $Q$, and are equal to $A^{^{2|R|\cdot k}}$, also in accordance with s.\ref{3.5}.

\subsection{Remaining mystery}

Thus, we see that the Gaussian distribution is a property of coefficients of
quantum numbers when they are re-expanded in powers of $z=q-q^{-1}$,
and, in an appropriate setting, this can de the technical reason for the Gaussianity of the
LMOV numbers.
However, at least three kinds of questions remain.

First, the very property of quantum numbers needs to be better understood.
Appearance of the golden section $\frac{1+\sqrt{5}}{2}$, which is the $q$-preimage
of $z=1$ gives rise to countless associations, and many of them deserve further
analysis.

Second, for this property to serve as an explanation for the Gaussianity, the knot polynomials should be somehow approximated by {\it isolated} quantum numbers.
If one adds or subtracts many adjacent quantum numbers, the Gaussianity breaks down:
one can play with the combinations like $\sum_{i=0}^s c_i[2K-2i+1]$
with large $K$ and small $s=2,3,4,5$  to see that the Gaussianity of $z$-coefficients
can be easily violated. The reason for this is a rather weak $K$-dependence of the
"amplitude" $I_K$ in (\ref{Gl2}): it is enough for $c_{i+1}/c_{i}$ to exceed
$\frac{\sqrt{5}+1}{\sqrt{5}-1} = 2.62\ldots$ to make the next term dominating
over the previous one. In the previous subsection, we demonstrated how this problem can be eliminated
by picking up the contributions with a given power of $A$ and by increasing parameters
like the representation size, the evolution parameter and the number of strands:
this often allows one to convert knot polynomials into combinations of
parametrical-separated quantum numbers (i.e. to forbid adjacent numbers).
This argument, however, is yet not established well enough to provide a convincing
explanation for all cases.

Third, the LMOV numbers count the holomorphic maps of Riemann surfaces,
and needed is also an interpretation of the Gaussian distribution in these terms.
This can get natural if there were independent windings around numerous cycles,
then, it should be demonstrated that this cycle-counting provides the right
numbers and reproduces the parameters of the Gaussian distributions ($\mu$, $\sigma$, $I$)
for various knots.

Each of these directions can be further split in many smaller questions,
which all deserve careful investigation.
What is important, now we have enough experimental information and appropriate
theoretical setting (statistical properties of quantum numbers after a Toda like
transform $q\to z$) to begin a real study of the LMOV numbers
and the structure of knot related topological theories.

\section{Conclusion}

From analysis of a large variety of the LMOV integers provided by recent
advance in the arborescent knot calculus \cite{arbor,evo3,MMMRS,MMf,Rama2},
we made an "experimental" observation leading to the following

\paragraph{Conjecture:} {\it Absolute values of the LMOV numbers
$N_{Q,g,n}^{\cal K}$ for big enough representations $Q$ approach a Gaussian/binomial
distribution in $g$ with just three  $Q,n$-dependent parameters.}
A similar bell-like behaviour is observed for the distribution in $n$.

\bigskip

This Gaussian-like asymptotics is achieved faster (in $|Q|$)
for "irregular" knots ${\cal K}$, while for knots with regular
sub-structures one can first (at small $Q$) obtain a superposition
of several Gaussian bells.

This observation opens a way to counting the {\it true fundamental}
degrees of freedom in the knot/link-associated Gopakumar-Ooguri-Vafa
topological theories, which can be a first step towards understanding of
what they really are.

Conceptual significance of the emergence and observation of {\it distributions}
in exactly-solvable Chern-Simons theory is still to be appreciated.
We just mention that it reflects the general expectations in non-linear algebra, see \cite{NLA}.
It would be quite interesting to relate the Gaussianity of the LMOV distributions with the attempt \cite{HoMan}
to statistically characterize the entire variety of colored HOMFLY-PT polynomials
in terms of the algebro-geometric approach to Mandelbrot-set theory \cite{DMmand, AndMmand}.
Another option is to make one step further and look at the distribution of
the three parameters of the Gaussian bell in the space of knots.
An experience from \cite{Vas} (in the case of Vassiliev invariants)
demonstrates that such ``landscape'' distributions can exhibit new (next-level)
mysterious structures.
In ss.\ref{3.5} and \ref{evo} we demonstrated a possibility and power of this approach,
its development is one of the obvious tasks for the future.

Technically, the Gaussian/binomial distributions in $z$-variable appear typical
for quantum numbers and their ratios,
and this is inherited by knot polynomials.
However, the real meaning of this phenomenon and its QFT/stringy interpretation
still needs to be understood much better.
What we demonstrated in this text is that after the achievements of \cite{MMM,Hcalc,arbor,evo3,MMMRS,MMf,Rama2},
the experimental data is now sufficient to check and develop any hypothesis
one can make about the LMOV numbers,
and this opens a way to full understanding of this intriguing branch of science,
a key one for developing a quantitative approach to the AdS/CFT correspondence
and other avatars of the open-close string duality.

\section*{Acknowledgements}

This work was funded by the Russian Science Foundation (Grant No.16-12-10344).

\end{document}